\documentclass[amsmath,nofootinbib,amssymb,prd]{revtex4}
\pdfoutput=1

\usepackage[utf8]{inputenc}
\usepackage{float}
\usepackage{epsfig}
\usepackage{graphicx}
\usepackage{color}
\usepackage{tabularx}
\usepackage{color}
\usepackage{verbatim}
\usepackage[inline, final]{showlabels}
\usepackage{tensor}
\usepackage{enumerate}

\usepackage{subcaption}
\usepackage{color}
\usepackage[bookmarks,colorlinks,linkcolor=webblue,citecolor=webgreen]{hyperref}	
\definecolor{webgreen}{rgb}{0, 0.5, 0} 
\definecolor{webblue}{rgb}{0, 0, 0.5} 
\definecolor{webred}{rgb}{0.5, 0, 0} 

\newcommand{\Od}[1]{\mathcal{O}\!\left( #1 \right)}
 
\newcommand{\ii}{\text{i}}
\newcommand{\di}{\text{d}}

\renewcommand{\v}[1]{\boldsymbol{#1}}

\begin{document}
\preprint{}
\title{Proca in the sky}
\author{Lavinia Heisenberg\footnote{lavinia.heisenberg@phys.ethz.ch}, Hector Villarrubia-Rojo\footnote{herojo@phys.ethz.ch}}
\affiliation{Institute for Theoretical Physics, ETH Z\"{u}rich, Wolfgang-Pauli-Strasse 27, 8093, Z\"{u}rich, Switzerland}
\date{\today}

\begin{abstract}
The standard model of cosmology, the $\Lambda$CDM model, describes the evolution of the 
Universe since the Big Bang with just a few parameters, six in its basic form. 
Despite being the simplest model, direct late-time measurements of the Hubble 
constant compared with the early-universe measurements result in the so-called $H_0$ tension. 
It is claimed that a late time resolution is predestined to fail 
when different cosmological probes are combined. In this work, we shake the ground 
of this belief with a very simple model. 
We show how, in the context of cubic vector Galileon models, the Hubble tension
can naturally be relieved using a combination of CMB, BAO and SNe observations \emph{without} 
using any prior on $H_0$. The tension can be reduced even further by including the local 
measurement of the Hubble constant.
\end{abstract}

\maketitle
\tableofcontents

\section{Introduction}
The standard model of cosmology rests on two fundamental pillars, the cosmological 
principle and General Relativity. The former states that the observable properties of the 
Universe are isotropic and homogeneous, which on the other hand enables to find exact 
solutions to Einstein's field equations. In the standard formulation of General Relativity 
(see \cite{BeltranJimenez:2019tjy,Heisenberg:2018vsk} for alternative formulations) the 
fundamental object is the symmetric $4\times4$ metric tensor, $g_{\mu\nu}$. The 
cosmological principle greatly simplifies the metric to be of the 
Friedmann-Lemaitre-Robertson-Walker form, where the dynamics given by Einstein's field 
equations are solely captured by the scale factor. Matter fields are represented by the 
homogenous and isotropic energy-momentum tensor in form of pressure and energy density. The 
governing equations can be brought into a single equation for the Hubble parameter 
$\mathcal{H}=\dot{a}/a$ as a function of dimensionless density parameters
\begin{equation}
\mathcal{H}^2=a^2 H_0^2\left( \Omega_{r}a^{-4}+\Omega_{m}a^{-3}+\Omega_{K}a^{-2}+\Omega_{\Lambda} \right)\,,
\end{equation}
with the density parameters of radiation $\Omega_{r}$, non-relativistic matter 
$\Omega_{m}$, curvature $\Omega_{K}$ and cosmological constant $\Omega_{\Lambda}$, and 
the Hubble constant $H_0$. Even though this simple $\Lambda$CDM model is in good agreement 
with cosmological observations, the Hubble tension has created a growing concern. The 
distance scale measurement of the Hubble constant based on Cepheids from the SH0ES 
collaboration gave a value $H_0=74.03\pm 1.42$ \cite{Riess:2019cxk}, which is more than 
4$\sigma$ away from the value inferred from Planck $H_0=67.44\pm 0.58$ \cite{Aghanim:2018eyx}. 
This discrepancy may be due to systematic errors, but it could also signal deviations 
from the $\Lambda$CDM model.
In this work, we will assume the latter. We will consider an extension of the standard 
model in the 
presence of an additional vector field, playing the role of dark energy. It is claimed that if the Hubble tension is 
resolved through modifications in the late time universe, this will be very difficult to 
reconcile with early universe measurements like BAO. Here, we show with a very simple 
model of a vector Galileon how this belief is easily circumvented.

Mostly studied extensions of the $\Lambda$CDM model are based on an additional scalar 
field. In the cosmological context, theories of the scalar Galileon and Horndeski 
\cite{Nicolis:2008in,Deffayet:2009wt,Horndeski:1974wa} have received quite some attention. 
Simple scalar models like Quintessence generically fail to address the Hubble tension 
\cite{Banerjee:2020xcn}. Cubic Horndeski and Galileon type of scalar theories typically 
fail to reconcile different cosmological measurement including the ISW-galaxy density
cross-correlations \cite{Renk:2017rzu}. On the other hand, as it was shown in 
\cite{deFelice:2017paw} for the first time, simple vector models can readily alleviate 
the Hubble tension due to a phantom-like behaviour of the background. The background 
analysis was further extended to linear perturbations in \cite{DeFelice:2020sdq}. Here, 
we present the results of our own implementation into a Boltzmann code based on \verb|CLASS|
\cite{Blas:2011rf}, using also \verb|MontePython| \cite{Audren:2012wb, Brinckmann:2018cvx} to 
constrain the cosmological parameters and \verb|GetDist| \cite{Lewis:2019xzd} to analyze the posteriors and
plot the results. We will consider a specific model of the cubic 
Generalized Proca (GP) theory. It represents a simple but rich subclass of Generalized Proca 
interactions \cite{Heisenberg:2014rta,Allys:2015sht,Jimenez:2016isa}. 

This paper is organized as follows. In Section \ref{sec:SimpleModel} we present the 
subclass of Generalized Proca models that we focus on. Section 
\ref{sec:Background} analyzes the cosmological behaviour of the GP model at the 
background level. Section \ref{sec:Perturbations} extends the analysis to include 
perturbations. In Section \ref{sec:Observations} we constrain the free parameters of
the model, performing a detailed comparison with observational data, and address the 
Hubble tension. Finally, Section \ref{sec:Conclusions} gathers the main results and 
discusses some prospects for future work.

\section{A simple Generalized Proca model}\label{sec:SimpleModel}
	Galilean interactions for a spin-1 field can only be constructed for the massive case. 
	These are the Generalized Proca theories 
	\cite{Heisenberg:2014rta,Allys:2015sht,Jimenez:2016isa,Heisenberg:2018vsk}. 
	They constitute the most general Lagrangians for a massive vector field with derivative 
	self-interactions, which nevertheless give rise to second order equations of motion 
	and propagate 3 physical modes. From the six possible Lagrangians we will consider 
	only a simple subclass, which also automatically satisfies the bounds on the 
	gravitational waves speed. The generalized Proca theories have opened up a promising 
	avenue for phenomenological applications in cosmology and black hole physics 
	\cite{Tasinato:2014eka,DeFelice:2016yws,DeFelice:2016uil,Chagoya:2017fyl,
	Heisenberg:2017hwb}. In this work, we are interested in its cosmological implications, 
	specially in the context of the Hubble tension. The action for the model that 
	we consider is
	\begin{equation}
		S = \int\di^4 x\,\sqrt{-g}\big\{G_4 R + \alpha F + G_2(X) + G_3(X)\nabla^\mu A_\mu\big\}\ ,
	\end{equation}
	with the short-cut notations introduced
	\begin{equation}
		F\equiv -\frac{1}{4}F^{\mu\nu}F_{\mu\nu}\ ,\qquad X\equiv -\frac{1}{2}A^\mu A_\mu\ ,
		\qquad G_4 \equiv \frac{1}{16\pi G}\ .
	\end{equation}
	The energy-momentum tensor, defined as
	\begin{equation}
		T_{\mu\nu} \equiv -\frac{2}{\sqrt{-g}}\frac{\delta S_\text{\tiny Proca}}{\delta g^{\mu\nu}}\ ,
	\end{equation}
	is given by the following
	\begin{align}\label{energymomA}
		T_{\mu\nu} &= 
			\alpha\left(F\indices{_\mu^\rho}F_{\nu\rho}+Fg_{\mu\nu}\right)
			+ g_{\mu\nu} G_2 + A_\mu A_\nu G_{2,X}\nonumber\\
		&\quad + G_{3,X}\bigg(A_\mu A_\nu\nabla_\rho A^\rho 
			- A^\rho\left(A_\mu \nabla_\nu A_\rho+A_\nu \nabla_\mu A_\rho\right)
			+ g_{\mu\nu} A^\rho A^\sigma\nabla_\rho A_\sigma\bigg)\ .
	\end{align}
	The equations of motion of the vector field can be obtained as
	\begin{equation}
		\mathcal{E}_\mu \equiv \frac{\delta S_\text{\tiny Proca}}{\delta A^\mu} = 0\ ,
	\end{equation}
	and we have
	\begin{equation}\label{eomA}
		\mathcal{E}_\mu = \alpha \nabla_\nu F\indices{^\nu_\mu} - A_\mu G_{2,X}
			- G_{3,X}\Big(A_\mu \nabla_\nu A^\nu - A_\nu\nabla_\mu A^\nu\Big)\ .
	\end{equation}
	In the following section we will adapt the theory to the symmetries of the cosmological background, specify further the model and discuss the background observables. 

\section{The background evolution}\label{sec:Background}
	For any given model its confrontation with cosmological observations is a crucial 
	ingredient in testing the underlying theory. A natural starting point is the study of 
	the background evolution. Using the distance-redshift relation from Supernovae,
	BAO measurements and the full CMB data from \emph{Planck}, the parameters of the model can successfully be constrained.
	We will now particularize the results of the covariant equations in the previous section to a FLRW metric
	\begin{equation}
		\di s^2 = a^2(\tau)(-\di\tau^2 + \di\v{x}^2)\ ,
	\end{equation}
	and a field configuration for the vector field
	\begin{equation}
		A^0\equiv \frac{1}{a}\phi(\tau)\ ,\qquad A^i = 0\ .
	\end{equation}
	The equation of motion of the vector field \eqref{eomA} simplifies to
		\begin{subequations}\label{eq:bg_eom}
		\begin{align}
			\mathcal{E}_0 &= \phi\big(a G_{2,X} + 3\phi\mathcal{H}G_{3,X}\big) = 0\ ,
				\label{eq:bg_eom0}\\
			\mathcal{E}_i &= 0\ .
				\label{eq:bg_eomi}
		\end{align}
		\end{subequations}
	The energy-momentum tensor \eqref{energymomA}, on the other hand, particularizes into
		\begin{subequations}\label{eq:bg_tmunu}
		\begin{align}
			T\indices{^0_0} &= G_2 - \phi^2G_{2,X} - 3a^{-1}\mathcal{H}\phi^3G_{3,X}\ ,
				\label{eq:bg_t00}\\
			T\indices{^0_i} &= 0\ ,
				\label{eq:bg_t0i}\\
			T\indices{^i_j} &= \delta\indices{^i_j}\big(G_2-a^{-1}\phi^2\dot{\phi}\,G_{3,X}\big)\ ,
				\label{eq:bg_tij}
		\end{align}
		\end{subequations}
	where $\dot{\phantom{a}}\equiv \di/\di \tau$.	
	The usual fluid variables for the Proca field are
	\begin{subequations}
	\begin{align}
		\rho_A &= - G_2\ ,\\
		P_A &= G_2 + \frac{1}{3\mathcal{H}}\dot{G}_2\ ,\\
		w_A &= \frac{P_A}{\rho_A} =-1 -\frac{\dot{G}_2}{3\mathcal{H}G_2}\ .
	\end{align}
	\end{subequations}		
	Deriving the constraint \eqref{eq:bg_eom0} with respect to $\tau$, we get 
	(for $\phi\neq 0$)
	\begin{equation}\label{eq:bg_constraint2}
		aG_{2,XX} = -3\phi\mathcal{H}G_{3,XX} - 3\phi^{-1}\mathcal{H}G_{3,X}
			-3\dot{\phi}^{-1}\left(\dot{\mathcal{H}}-\mathcal{H}^2\right)G_{3,X}\ .
	\end{equation}	
	The Einstein equations yield the usual Friedmann equations plus the contribution from our
	dark energy fluid
	\begin{align}
		\mathcal{H}^2 &= \frac{8\pi G a^2}{3}\rho + \frac{8\pi G a^2}{3}\rho_A\ ,
			\label{eq:bg_friedmann1}\\
		\dot{\mathcal{H}}-\mathcal{H}^2 &= -4\pi Ga^2(\rho+P) 
			- 4\pi G a^2 (\rho_A+P_A)\ .
			\label{eq:bg_friedmann2}
	\end{align}
	where $\rho$ and $P$ stand for the density and pressure of all the components except the
	Proca field.
	
	\subsection{Dark energy model}
Following the studies in \cite{DeFelice:2016yws,DeFelice:2016uil} we will consider a promising dark energy model, where the
general functions are chosen to be polynomials
		\begin{equation}\label{eq:bg_polynomial}
			G_2(X) = b_2X^{p_2}\ ,\qquad G_3(X) = b_3 X^{p_3}\ ,
		\end{equation}
		which yield from the constraint equation \eqref{eq:bg_eom0}
		\begin{equation}\label{eq:bg_defphi}
			\phi = \phi_0\left(\frac{\mathcal{H}}{aH_0}\right)^{-1/p}\ ,
		\end{equation}
		where
		\begin{equation}
			p\equiv 1-2(p_2-p_3)\ ,\qquad \phi_0 = \left(-\frac{2^{p_3}b_2p_2}{3H_02^{p_2}b_3p_3}\right)^{1/p}\ .
		\end{equation}
		With the previous relations, the evolution of the density follows
		\begin{equation}\label{eq:bg_G2dot}
			\dot{\rho}_A = -\frac{2s}{\mathcal{H}}\left(\dot{\mathcal{H}}-\mathcal{H}^2\right)\rho_A\ ,
				\qquad s\equiv \frac{p_2}{p}\ .
		\end{equation}
		Using the Friedmann equations, it can also be rewritten as
		\begin{equation}
			\dot{\rho}_A = \frac{3\mathcal{H}(\rho + P)s\,\rho_A}{\rho + (1+s)\rho_A}\ ,
		\end{equation}
		where the equation of state in this case is
		\begin{align}\label{eq:bg_wA}
			w_A &= -1 - \frac{(1+w)s}{1+(1+s)\rho_A/\rho}\;.
		\end{align}
		Finally, writing $\rho_\text{tot}=\rho+\rho_A$, we can reduce the differential equation to an
		algebraic one
		\begin{equation}\label{eq:bg_algebraic_solution}
			-\frac{1}{s\rho_A}\frac{\di\rho_A}{\di\log a} 
				= \frac{1}{\rho_\text{tot}}\frac{\di\rho_{\text{tot}}}{\di\log a}\qquad\to\qquad
			\rho_A\propto \rho_\text{tot}^{-s} = (\rho+\rho_A)^{-s}\ .
		\end{equation}
		For later reference, we introduce the following short-hand notation
		\begin{equation}\label{eq:bg_defR}
			\mathcal{R}\equiv \frac{\rho_A}{\rho_\text{tot}}\ ,\qquad
			\tilde{\mathcal{R}}\equiv 1-\mathcal{R}= \frac{\rho}{\rho_\text{tot}}\ ,\qquad
			\mathcal{X}\equiv -\frac{1+w_A}{s} = \frac{(1+w)\tilde{\mathcal{R}}}{1+s\mathcal{R}}\ .
		\end{equation}
		Rewriting \eqref{eq:bg_G2dot} in terms of these variables we have
		\begin{equation}
			\dot{\mathcal{R}} = \frac{3\mathcal{H}(s+1)(1+w)}{1+s\mathcal{R}}\mathcal{R}(1-\mathcal{R})\ .
		\end{equation}
		From this last expression we can see that, at late times, the Proca field evolves 
		toward the de Sitter attractor $\mathcal{R}\to 1$ and $w_A\to -1$.
	
	\subsection{Physical effects}
		The concrete dark energy model that we have considered in \eqref{eq:bg_polynomial} 
		contains in principle four free parameters $(b_2, b_3, p_2, p_3)$. However, only two
		combinations of these parameters produce effects on the background. Assuming a
		flat universe, the value of the dark energy density today is fixed, thus fixing one of
		these combinations. In this case, all the modifications to the background evolution
		are governed by a single parameter $s$, defined in \eqref{eq:bg_G2dot}.
		
\begin{figure}[h!]
			\centering
			\begin{subfigure}[t]{0.48\textwidth}
				\includegraphics[scale=0.55]{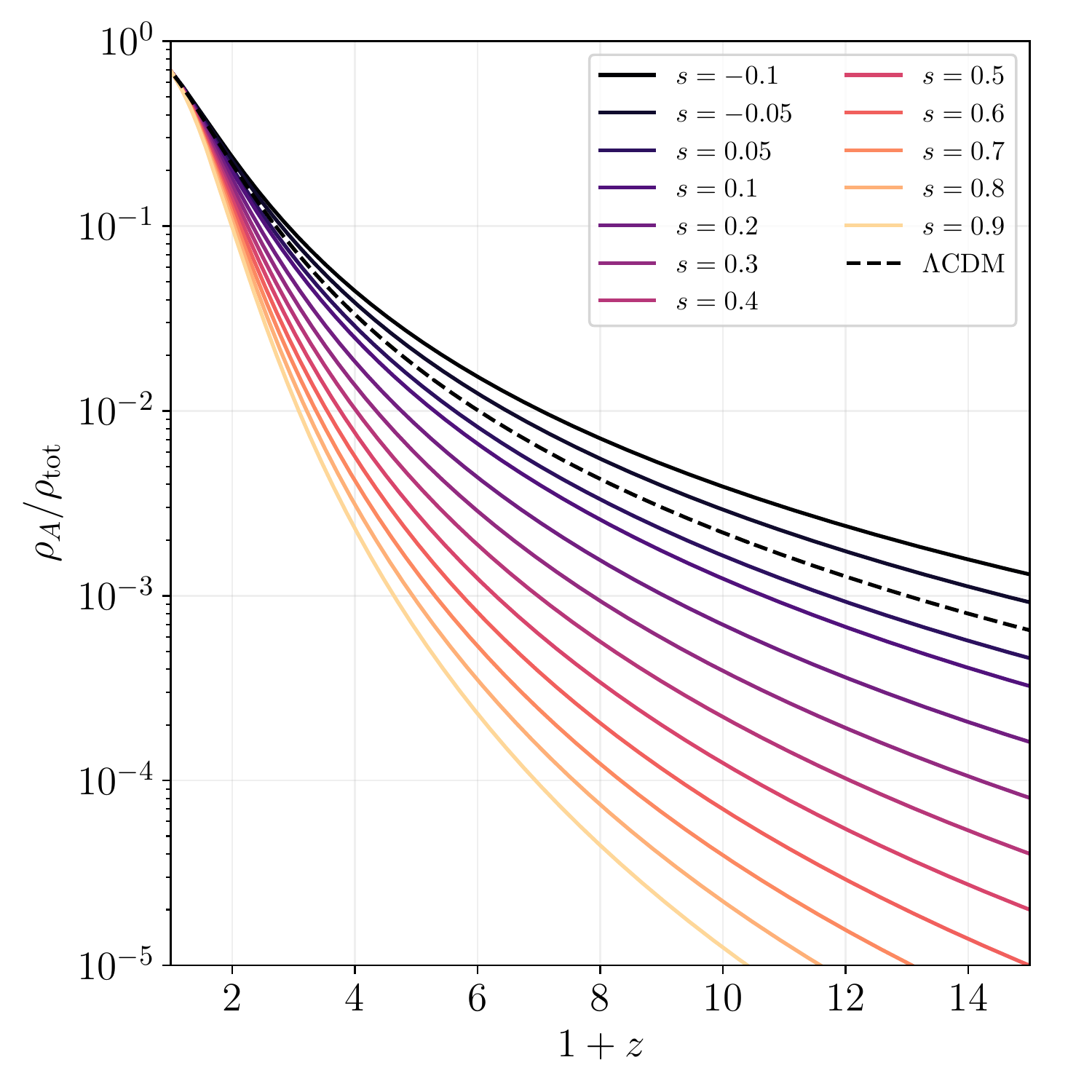}
			\end{subfigure}
			\begin{subfigure}[t]{0.48\textwidth}
				\includegraphics[scale=0.55]{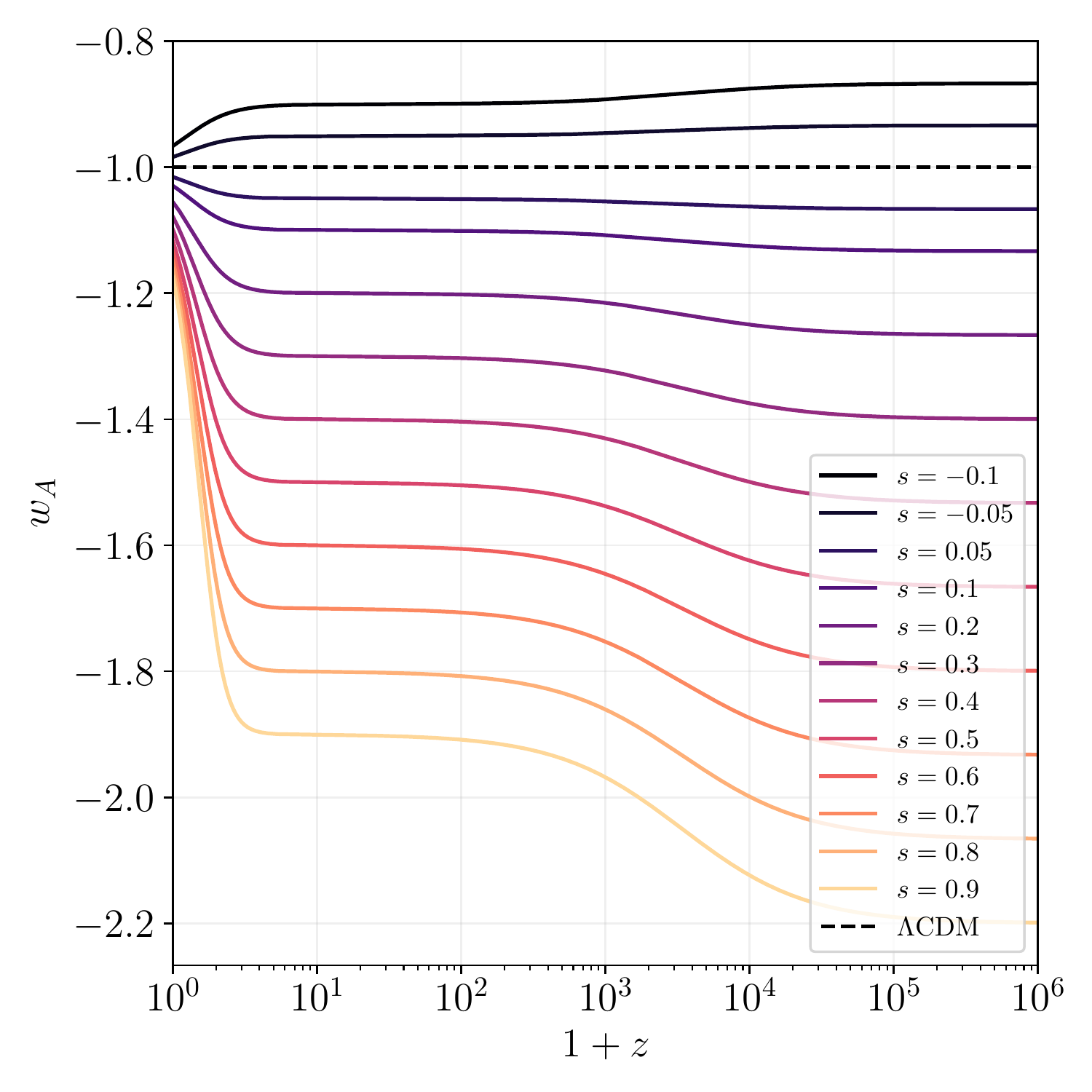}
			\end{subfigure}
			\caption{(Left) Energy density of the Proca field normalized
				to the total density. (Right) Equation of state for different values
				of $s$. Notice that the equation of state is phantom for $s>0$. In both
				plots, the dotted line represents a cosmological constant. }
			\label{fig:density_w}
	    \end{figure}

\begin{figure}[h!]
			\centering
			\begin{subfigure}[t]{0.48\textwidth}
				\includegraphics[scale=0.55]{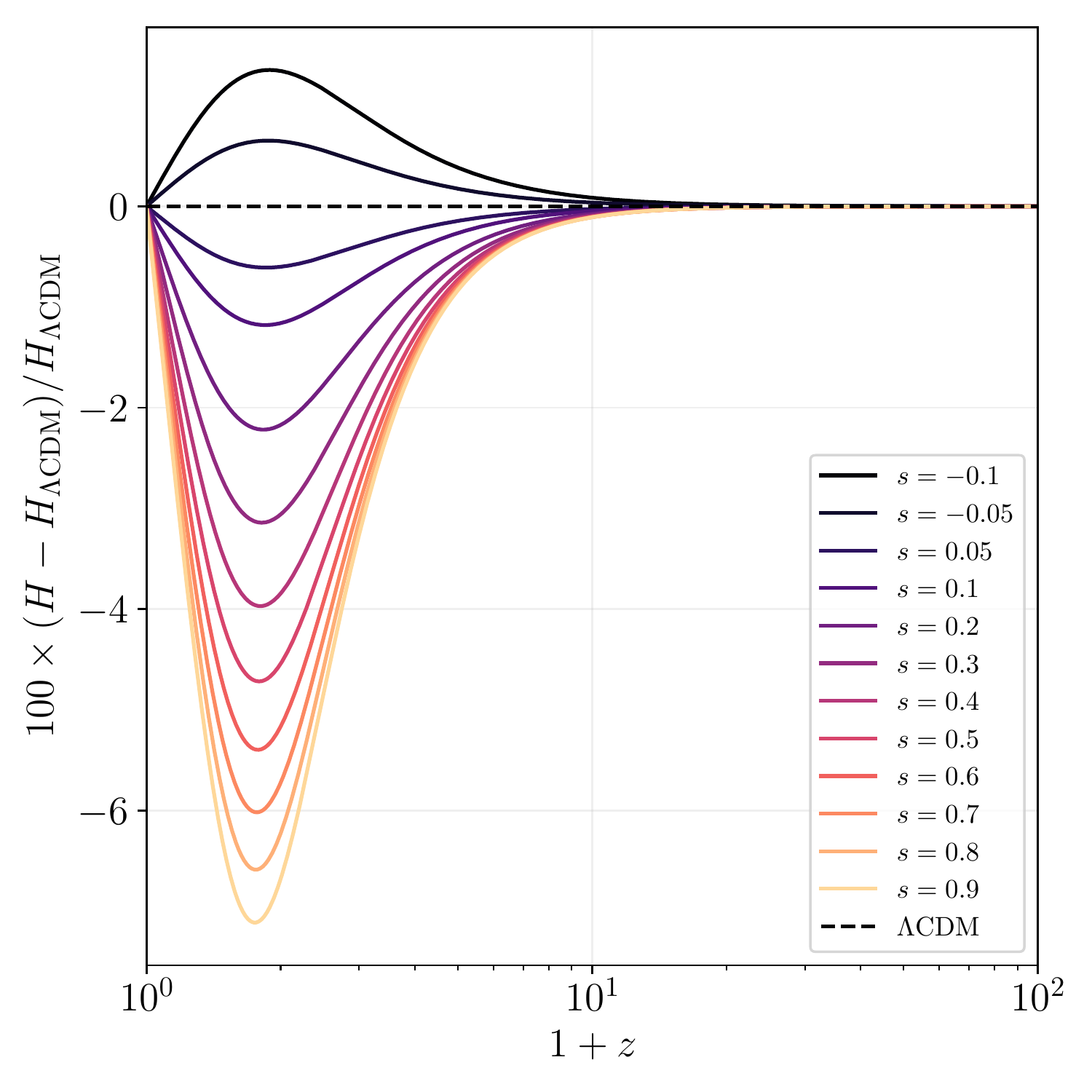}
			\end{subfigure}
			\begin{subfigure}[t]{0.48\textwidth}
				\includegraphics[scale=0.55]{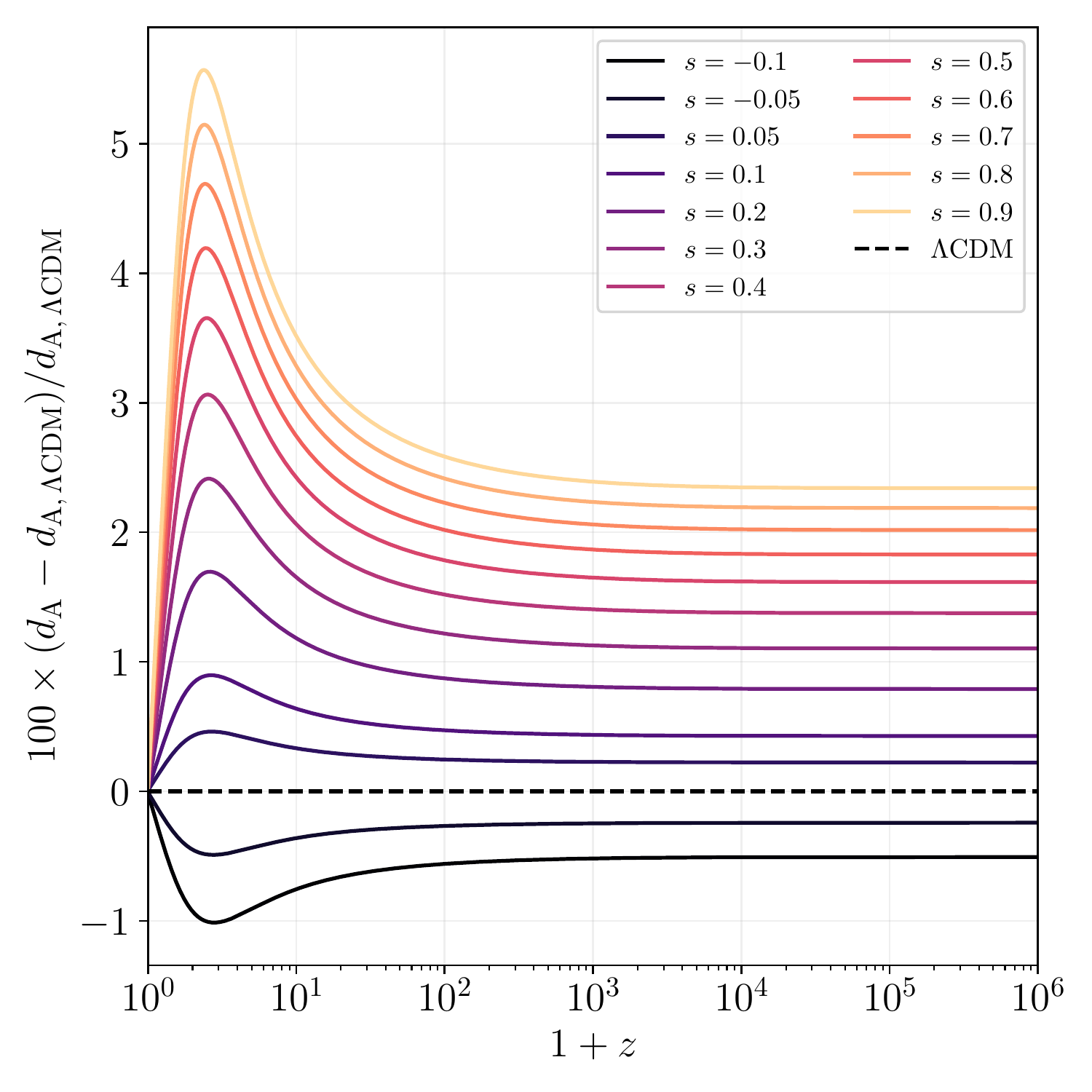}
			\end{subfigure}
			\caption{Relative deviations in the Hubble parameter (Left) and the 
				angular diameter distance (Right) with respect to the $\Lambda$CDM 
				model.}
			\label{fig:H_dA}			
	    \end{figure}

		This parameter determines the evolution of the dark energy fluid in \eqref{eq:bg_wA}.
		At early times, the dark energy fluid is subdominant and its equation of state can
		be directly related to the equation of state of the matter-radiation fluid
		\begin{equation}
			1+w_A \simeq -(1+w)s\ ,\qquad \rho_A\ll\rho\ .			
		\end{equation}
		In particular, for $s=-1$ the fluid tracks exactly the evolution of the dominant
		component at the time. On the other hand, at late times, the fluid approaches the
		cosmological constant behaviour
		\begin{equation}
			w_A\simeq -1\ ,\qquad \rho_A\gg \rho\ .
		\end{equation}
		It is important to notice that for $s>0$, that will be our main concern, the
		dark energy is phantom-like, i.e. $w_A<-1$. Figure \ref{fig:density_w} contains
		the evolution of the energy density and the equation of state for different values
		of $s$. These modifications have a deep impact in the late-time evolution of
		the Hubble constant, as shown in Figure \ref{fig:H_dA}. 

		Modifications to the late-time expansion history can be directly tested with
		observations of (uncalibrated) SNe Ia. In Section \ref{sec:Observations}, we will 
		use these observations, among other probes, to constrain the model. 
		Changes in $H(z)$ also affect the evolution of the perturbations and leave an 
		imprint in the CMB and in the matter distribution. We will analyze these effects
		in the next section, after studying the evolution of the perturbations to the 
		Proca field.
        
\section{Perturbations}\label{sec:Perturbations}
	In this section we consider a perturbed FLRW metric, taking only into account scalar
	perturbations
	\begin{equation}
		\di s^2 = a^2(\tau)\Big(-(1+2\Psi)\di\tau^2 + 2\partial_iB\,\di\tau\di x^i 
			+ (\delta_{ij} - 2\Phi\delta_{ij} - \partial_i\partial_jE)\di x^i\di x^j\Big)\ .
	\end{equation}
	Similarly, the scalar perturbations to the Proca field are
	\begin{subequations}
	\begin{align}
		A^0 &= \frac{1}{a}(\phi + \delta\phi)\ ,\\
		A^i &= a^{-2}\delta^{ij}\partial_j\chi_V\ .
	\end{align}
	\end{subequations}
	In Appendix \ref{app:gen_pert} we analyze the effect of a generic perturbation, 
	proving that the vector modes can be neglected. \\
	
	After plugging the Ansatz with perturbations into the vector field equations, one obtains		
		\begin{align}
			\mathcal{E}_0 &= \alpha\frac{k^2}{a^2}\bigg\{a(\delta\phi+2 \Psi\phi) 
				+\partial_\tau\left(\chi_V+a\phi B\right)\bigg\}
				+ \frac{a\phi\left(\dot{\mathcal{H}}-\mathcal{H}^2\right)}{\dot{\phi}\mathcal{H}}
				G_{2,X}\left(\delta\phi+\Psi\phi\right)\nonumber\\
			&\quad +a \Psi G_{2,X}\phi + \frac{G_{2,X}}{3\mathcal{H}}\left(k^2\chi_V
				+a\phi(3\dot{\Phi}+k^2\dot{E})\right)=0\ ,\\
			\frac{\ii k^i}{k^2}\mathcal{E}_i &= 
			\frac{\alpha}{a^2}\bigg\{\partial_\tau\big(a(\delta\phi+2\Psi\phi)\big)
					+\partial_\tau^2\big(\chi_V+ a\phi B\big)\bigg\}\nonumber\\
			&\quad -\frac{aG_{2,X}}{3\mathcal{H}}\left(\delta\phi +\Psi\phi \right)
				- \frac{G_{2,X}}{3\phi\mathcal{H}}\dot{\phi}\big(\chi_V+ a\phi B\big)=0\ .
		\end{align}	
		
	In a similar way, the energy-momentum tensor becomes this time
		\begin{align}
			\delta T\indices{^0_0} &= \phi G_{2,X}\left(\delta\phi+\Psi\phi\right)
					-G_{2,X}\frac{\phi^2}{\mathcal{H}\dot{\phi}}\left(\dot{\mathcal{H}}-\mathcal{H}^2\right)
					\left(\delta\phi+\Psi\phi\right) - G_{2,X}\Psi\phi^2\nonumber\\
				&\quad -\frac{\phi}{3a\mathcal{H}}G_{2,X}\left(k^2\chi_V+a\phi(3\dot{\Phi}+k^2\dot{E})\right)\ ,\\
			\ii k^i\delta T\indices{^0_i} &= k^2\,\frac{\phi G_{2,X}}{3\mathcal{H}}\left(\delta\phi+\Psi\phi\right)\ ,\\
			\delta T\indices{^i_j} &= a^{-4}\delta\indices{^i_j}\partial_\tau\left\{
				\frac{a^4\phi}{3\mathcal{H}}G_{2,X}\left(\delta\phi + \Psi\phi\right)\right\}
				-\frac{1}{3\mathcal{H}}\phi\dot{\phi}\,G_{2,X}\Psi\delta\indices{^i_j}\ .
		\end{align}
		We will adopt the usual definitions for the sources of the Einstein equations
		\begin{equation}
			\delta T\indices{^0_0} \equiv -\delta\rho\ ,\qquad 
			\ii k^i \delta T\indices{^0_i} \equiv (\rho_A+P_A)\theta_A\ ,\qquad
			\delta T\indices{^i_j}\equiv \delta\indices{^i_j}\delta P\ .
		\end{equation}		
	The system of equations can be greatly simplified by using the following set of
	dimensionless variables
	\begin{subequations}
	\begin{align}
		\delta_\chi &\equiv \frac{k(\chi_V+a\phi B)}{a\phi}\ ,\\
		\delta_\phi &\equiv \frac{k(\delta\phi + \Psi\phi)}{\dot{\phi}}\ ,\\
		\mathcal{Z} &\equiv -\frac{k^2\alpha\phi}{a^3\rho_A}
			\Big(a\left(\delta\phi + 2\Psi\phi\right) + \partial_\tau\left(\chi_V+a\phi B\right)\Big)\ .
			\label{eq:redef_defZ}
	\end{align}
	\end{subequations}
	With these definitions, the perturbed fluid variables are
	\begin{subequations}
	\begin{align}
		\delta\rho_A + \frac{3\mathcal{H}}{k^2}(\rho_A+P_A)\theta_A &= 
			-\frac{\dot{\rho}_A}{\mathcal{H}}\frac{\phi}{\dot{\phi}}\left\{
				\left(\dot{\mathcal{H}}-\mathcal{H}^2\right)\frac{\delta_\phi}{k}
				+\mathcal{H}\Psi
				+\frac{1}{3}\left(k\delta_\chi - k^2(B-\dot{E})+3\dot{\Phi}\right)\right\}\ ,
				\label{eq:redef_rhoplustheta}\\
		(\rho_A+P_A)\theta_A &= k(1+w_A)\rho_A\delta_\phi\ ,
				\label{eq:redef_theta}\\
		\delta P_A &= a^{-4}\partial_\tau\left(\frac{a^4}{k^2}(\rho_A+P_A)\theta_A\right)
			- \frac{\dot{\rho}_A}{3\mathcal{H}}\Psi\ .
				\label{eq:reder_deltaP}
	\end{align}
	\end{subequations}
	On the other hand, the equations of motion are
	\begin{align}
		\mathcal{E}_0 &= \frac{a}{\phi}\left\{-\rho_A\mathcal{Z}
			+\delta\rho_A + \frac{3\mathcal{H}}{k^2}(\rho_A+P_A)\theta_A\right\}=0\ ,
			\label{eq:redef_E0}\\
		\frac{\ii k^i}{k^2}\mathcal{E}_i &= \frac{a\rho_A}{k^2\phi}\left\{
			-\dot{\mathcal{Z}}+\left(\frac{\dot{\phi}}{\phi}-\frac{\dot{\rho}_A}{\rho_A}-3\mathcal{H}\right)\mathcal{Z}
			+\frac{k}{3\mathcal{H}}\frac{\dot{\rho}_A}{\rho_A}(\delta_\chi+\delta_\phi)\right\}=0\ .
			\label{eq:redef_Ei}
	\end{align}
	The variable $\delta_\phi$ can be obtained from \eqref{eq:redef_rhoplustheta} and
	\eqref{eq:redef_E0}
	\begin{equation}\label{eq:pert_deltaphi1}
		\left(\dot{\mathcal{H}}-\mathcal{H}^2\right)\delta_\phi
			= -k\mathcal{H}\frac{\dot{\phi}}{\phi}\frac{\rho_A}{\dot{\rho}_A}\mathcal{Z}
				+k\left(\frac{k^2}{3}(B-\dot{E}) - \dot{\Phi}-\mathcal{H}\Psi\right)
				-\frac{k^2}{3}\delta_\chi\ .
	\end{equation}
	We can write a system for the two dynamical variables $\delta_\chi$ and $\mathcal{Z}$
	using the definition \eqref{eq:redef_defZ} and the equation of motion \eqref{eq:redef_Ei}
	\begin{align}
		\dot{\delta}_\chi &= -\left(\mathcal{H}+\frac{\dot{\phi}}{\phi}\right)\delta_\chi 
				-\frac{a^2\rho_A}{\alpha\phi^2 k}\mathcal{Z}
				-\frac{\dot{\phi}}{\phi}\delta_\phi -k\Psi\ ,
			\label{eq:pert_deltachi1}\\
		\dot{\mathcal{Z}} &= \left(\frac{\dot{\phi}}{\phi}-\frac{\dot{\rho}_A}{\rho_A}-3\mathcal{H}\right)\mathcal{Z}
				+\frac{k}{3\mathcal{H}}\frac{\dot{\rho}_A}{\rho_A}(\delta_\chi+\delta_\phi)\ .
			\label{eq:pert_Z1}
	\end{align}
	
	\subsection{Dark energy model}\label{sec:Perturbations_Polynomial}
		We can particularize the results to the dark energy model that we considered in
		\eqref{eq:bg_polynomial}. From now on we will work in the Newtonian gauge, setting
		$B=E=0$. We will also set the normalization $\alpha=1$ and measure
		the scalar field in Planck units $M^{-1}_\text{P}=\sqrt{8\pi G}$.  Finally, we 
		introduce one more variable redefinition, that is closely related
		to the velocity perturbation,
		\begin{equation}
			\mathcal{Q} \equiv -\frac{1}{p}\left(\mathcal{Z}+\frac{2ksp}{3\mathcal{H}}\delta_\chi\right)\ .
		\end{equation}
		The constraint \eqref{eq:pert_deltaphi1} in terms of the new variable is
		\begin{equation}
			(1+w_A)\delta_\phi = \frac{k}{3\mathcal{H}}\mathcal{Q} - \frac{2sk}{3\mathcal{H}^2}\left(\dot{\Phi}+\mathcal{H}\Psi\right)\ .
		\end{equation}
		After solving for the constraint, the system of equations for the evolution of 
		the perturbations is reduced to 
		\begin{subequations}\label{eq:pert_eoms}
		\begin{align}
			\dot{\mathcal{Q}} &= -2\mathcal{H}\left(1-\frac{3}{4}\mathcal{X}\right)\mathcal{Q}
				+\mathcal{H}\Big(s\mathcal{R}+3(1+s\mathcal{R})c_A^2\Big)\mathcal{Z}
				+ \frac{2k^2s}{3\mathcal{H}}\Psi\ ,
				\label{eq:pert_eom1}\\
			\dot{\mathcal{Z}} &= 3\mathcal{H}w_A\mathcal{Z}
				- \mathcal{H}\left(\frac{k^2}{3\mathcal{H}^2}+\frac{3}{2}\mathcal{X}\right)\mathcal{Q}
				+ \frac{2k^2 s}{3\mathcal{H}^2}\left(\dot{\Phi}+\mathcal{H}\Psi\right)\ .
				\label{eq:pert_eom2}
		\end{align}
		\end{subequations}
		where $\mathcal{R}$, $\mathcal{X}$ have been defined in \eqref{eq:bg_defR} and
		we have introduced 
		\begin{equation}\label{eq:pert_c2A_def}
			c^2_{\scriptsize A} = \left(1+s\mathcal{R}\right)^{-1}p^{-1}\left\{\frac{2sp\mathcal{R}}{3\phi^2}
				+\frac{1}{3}\left(1-sp\mathcal{R}\right) + \frac{1}{2}\left(1+2s-\frac{1}{p}\right)\mathcal{X}\right\}\ .
		\end{equation}
		The sources of the Einstein equations are
		\begin{align}
			(\rho_A+P_A)\theta_A &= \left(1+s\mathcal{R}\right)^{-1}\left\{
				\frac{k^2}{3\mathcal{H}}\rho_A\mathcal{Q}-s\mathcal{R}(\rho+P)\theta\right\}\ ,
				\label{eq:pert_src1}\\
			\delta\rho_A &= \rho_A\mathcal{Z} - \frac{3\mathcal{H}}{k^2}(\rho_A+P_A)\theta_A\ .
				\label{eq:pert_src2}
		\end{align}
		Again, $(\rho+P)\theta$ represents the momentum perturbation of all the fluids 
		except the Proca field. As we can see, the perturbations $\mathcal{Z}$ and 
		$\mathcal{Q}$ are directly related to the density and velocity perturbations
		of the dark energy fluid.
		
		When solving for the evolution of perturbations we will restrict the parameter 
		space, considering only models that satisfy the following consistency conditions.
		\begin{itemize}
			\item \emph{Absence of ghost and gradient instabilities}. It was shown in
				\cite{deFelice:2017paw} that these conditions amount to imposing
				\begin{align}
					Q_A &= \frac{3sp^2(1+s\mathcal{R})\mathcal{H}^2\mathcal{R}}{(1-sp\mathcal{R})^2a^2\phi^2}> 0\ ,\\
					c_A^2 &> 0\ ,
				\end{align}
				where $c_A^2$ defined in \eqref{eq:pert_c2A_def} corresponds to $c^2_S$ in 
				\cite{deFelice:2017paw} and $Q_A$ corresponds to $Q_S/a^3$ in the 
				same reference. The first condition can be simply imposed by restricting to 
				models where $s>0$.
			\item \emph{Absence of strong coupling}. Using the scaling relations \eqref{eq:bg_defphi} 
				and \eqref{eq:bg_algebraic_solution} one can show that, in the $\mathcal{R}\to 0$ limit, 
				\begin{equation}
					Q_A\sim \frac{\mathcal{H}^2\mathcal{R}}{a^2\phi^2}\propto \mathcal{R}^{\frac{sp-1}{p(s+1)}}\ .
				\end{equation}
				Since in the asymptotic past the dark energy fluid is subdominant, 
				$\mathcal{R}\to 0$, in order to avoid strong coupling problems, $Q_A\to 0$,
				we must impose
				\begin{equation}
					sp=p_2 < 1\ .
				\end{equation}
		\end{itemize}
		
	\subsection{Physical effects}\label{sec:Perturbations_PhysicalEffects}
		The equations of motion \eqref{eq:pert_eom1} and \eqref{eq:pert_eom2}, together
		with the sources \eqref{eq:pert_src1} and \eqref{eq:pert_src2}, determine the 
		evolution of the dark energy perturbations. In addition to the parameter $s$ that
		determines the background evolution, the perturbations also depend on the 
		parameters $p$, or $p_2=sp$, and $\phi_0$.
		 						
		\begin{figure}[ht]
			\centering
			\begin{subfigure}[t]{0.48\textwidth}
				\includegraphics[scale=0.55]{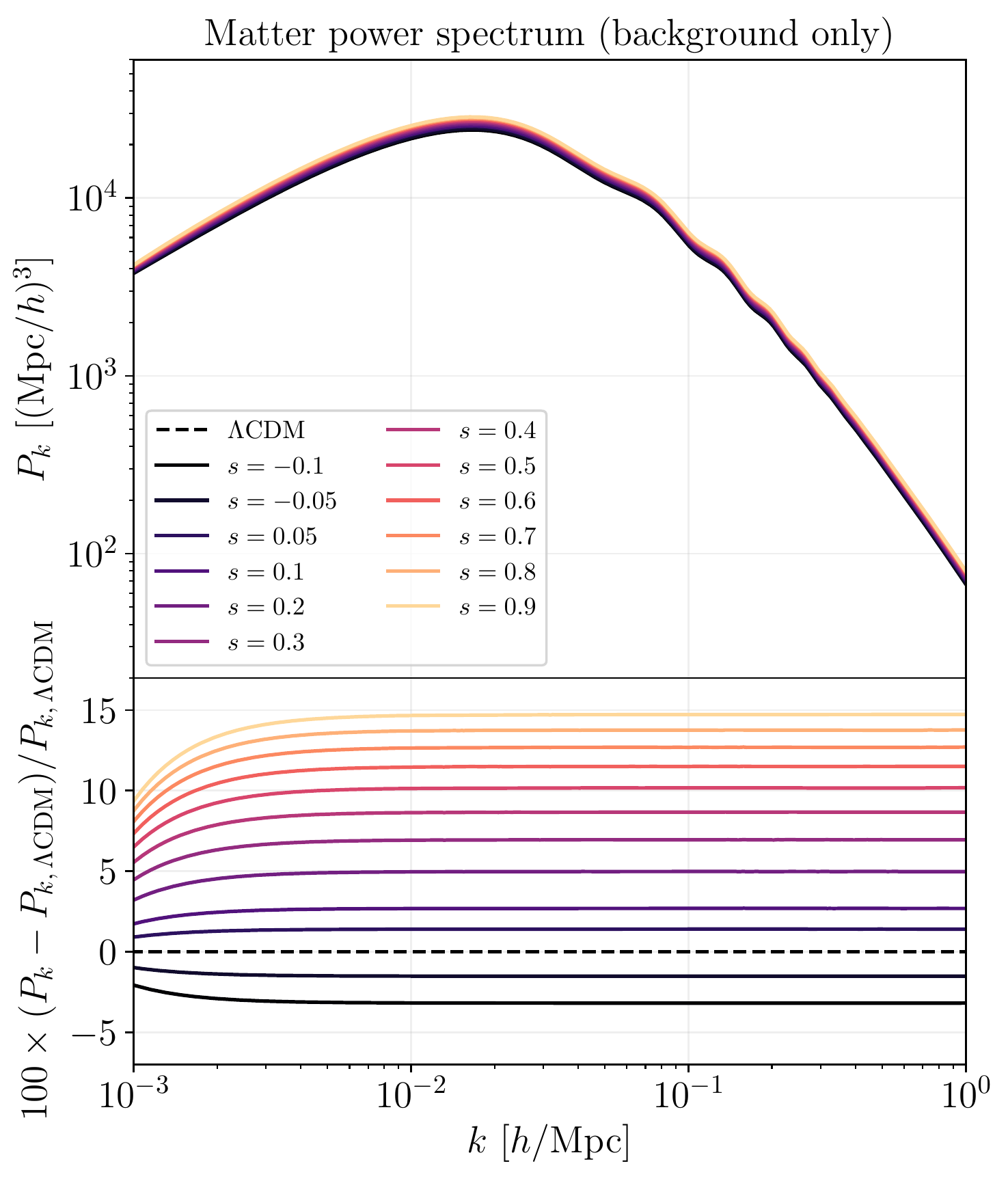}
			\end{subfigure}
			\begin{subfigure}[t]{0.48\textwidth}
				\includegraphics[scale=0.55]{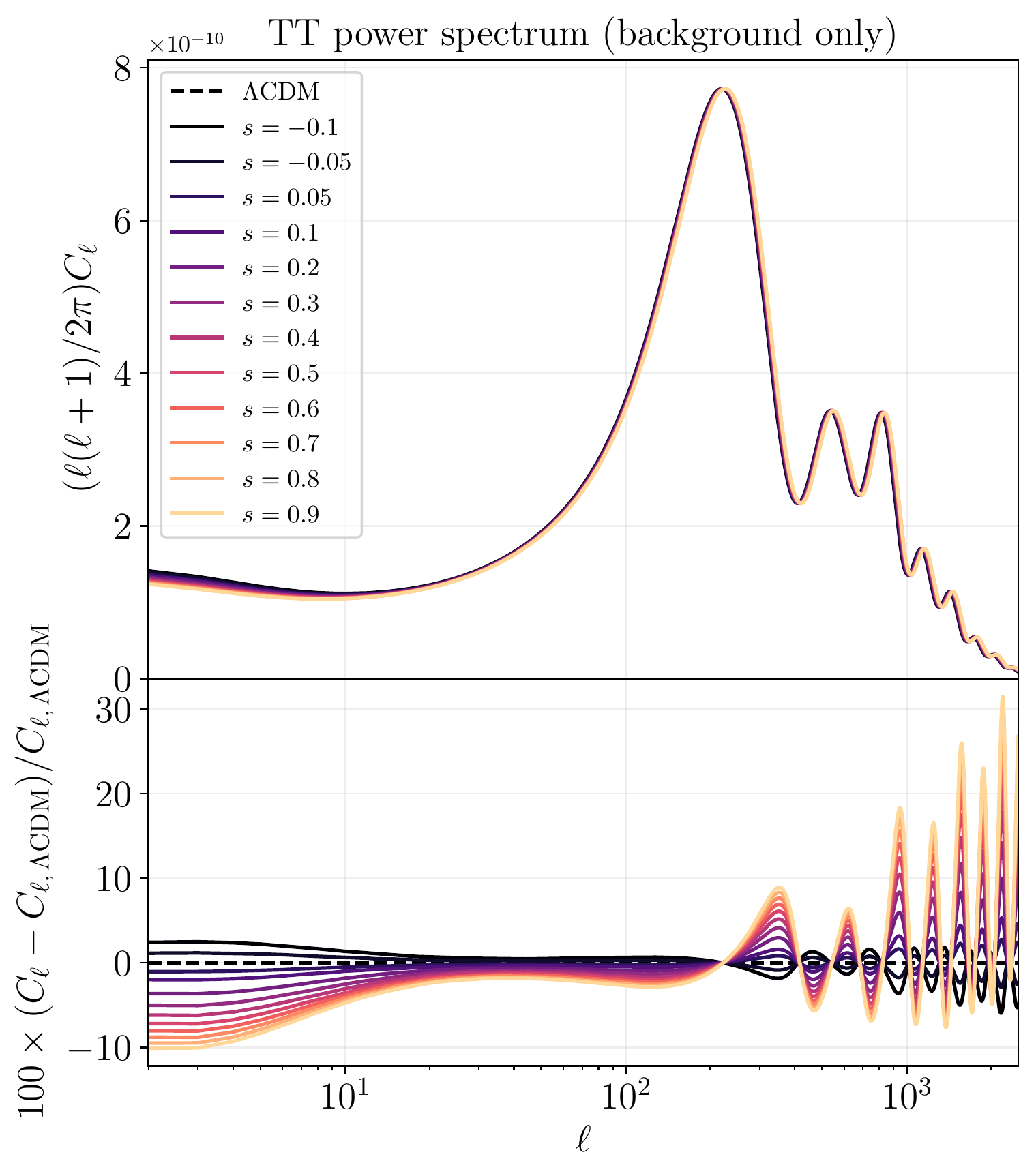}
			\end{subfigure}
			\caption{Matter and temperature power spectra, including only the modifications
				to the background discussed in Section \ref{sec:Background}. In the lower
				pannels we also show the relative deviations with respect to the $\Lambda$CDM
				result, depicted with a dotted line. As discussed in the main text, the
				modification of the expansion history has two key effects: (I) it changes 
				the angular diameter distance to decoupling, shifting the CMB peaks, (II)
				it changes the growth factor, modifying the $P_k$ amplitude and the CMB
				plateau.}
			\label{fig:Pk_Cl_bgonly}
	    \end{figure}
	    
	     \begin{figure}[ht]
			\centering
			\begin{subfigure}[t]{0.48\textwidth}
				\includegraphics[scale=0.55]{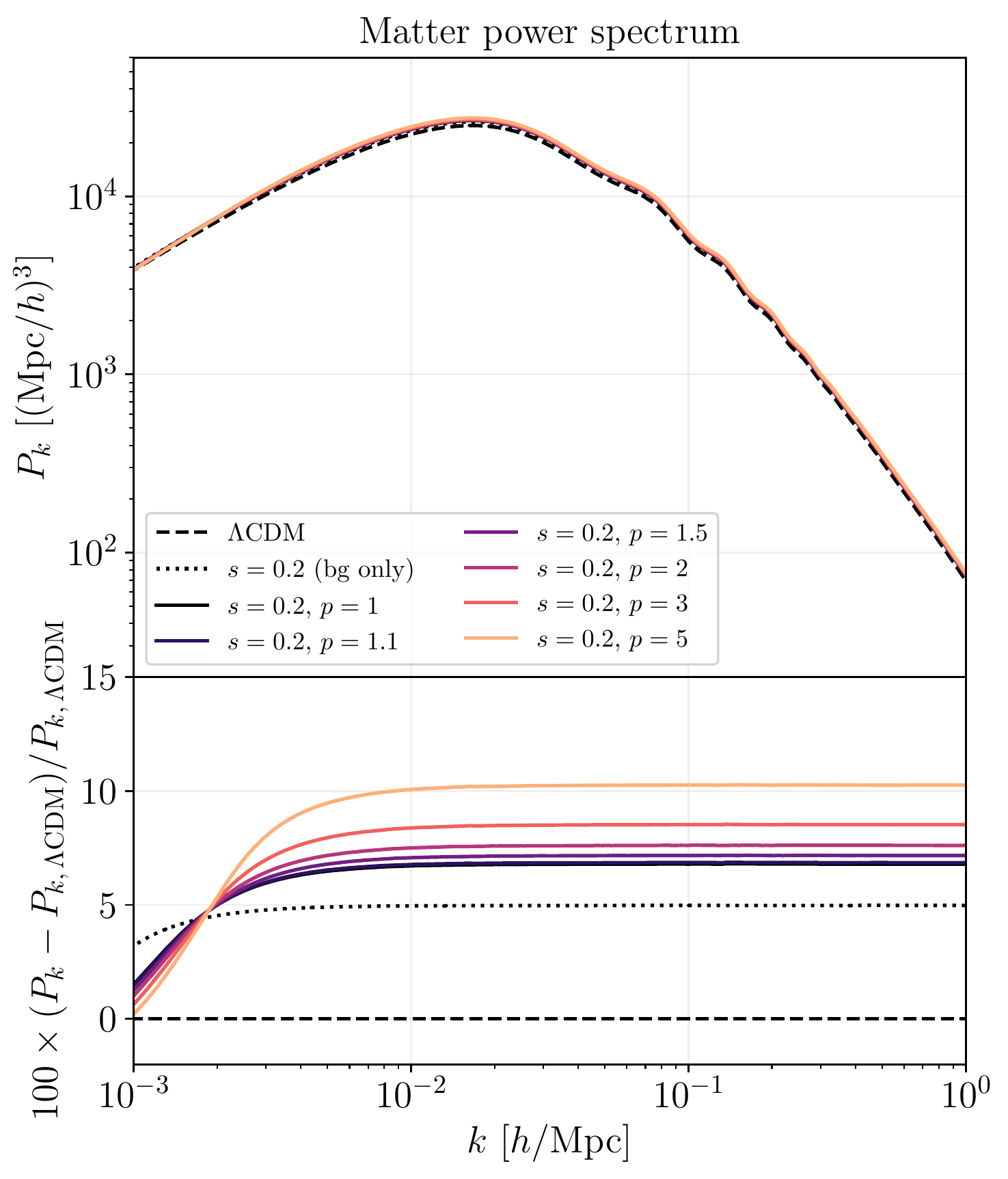}
			\end{subfigure}
			\begin{subfigure}[t]{0.48\textwidth}
				\includegraphics[scale=0.55]{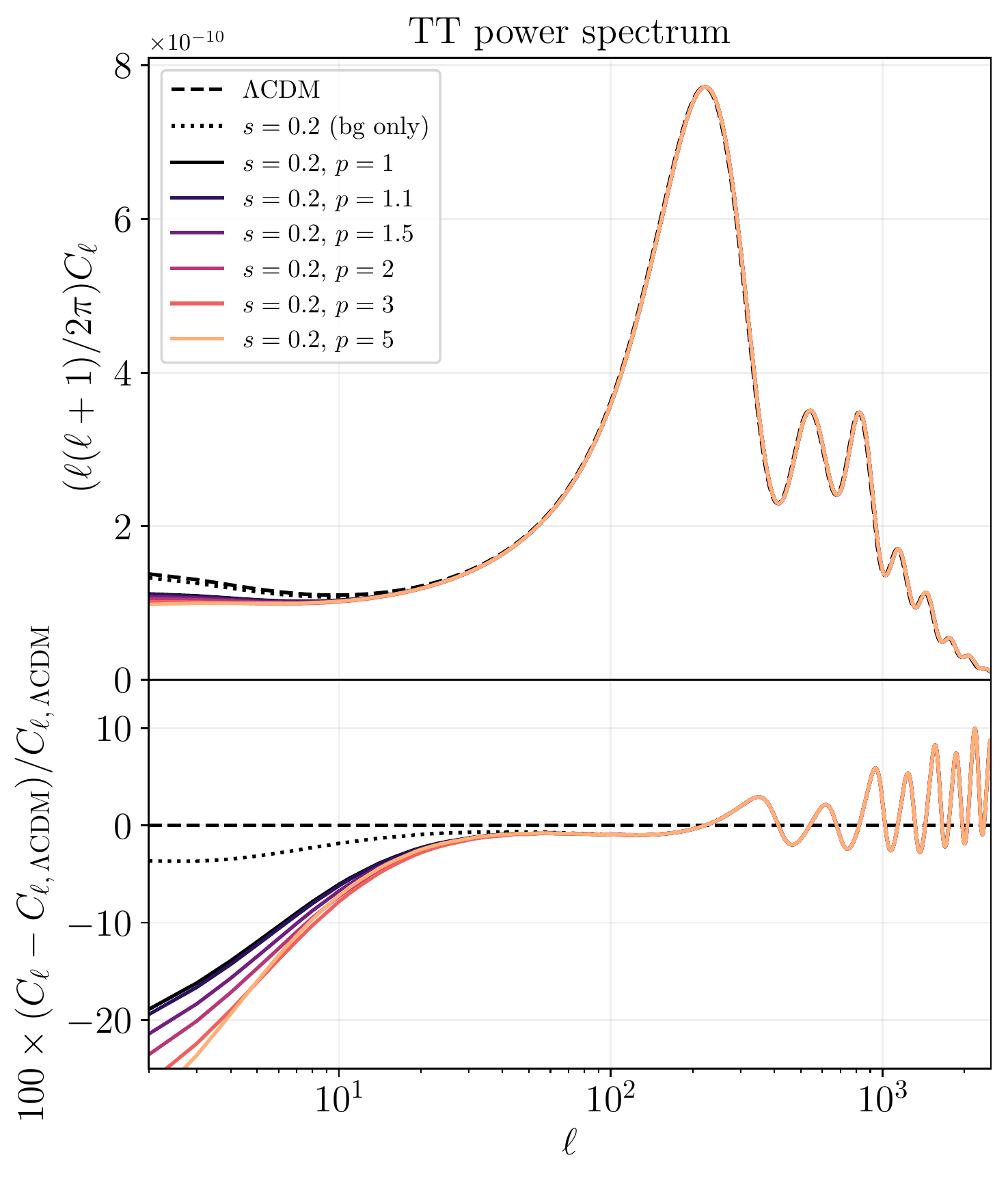}
			\end{subfigure}
			\caption{Matter and temperature power spectra, including both the modified
			background and the perturbations. As discussed in the text, the effects 
			observed can be traced back to the clustering properties of the dark energy 
			fluid, that lead to a late-time evolution of the metric potentials.}
			\label{fig:Pk_Cl}
	    \end{figure}	
	
		In order to disentangle all the physical effects, we have studied two GP
		models. The first one contains only modifications at the background level and
		the results are represented in Figure \ref{fig:Pk_Cl_bgonly}. The second one
		contains both the effects on the background and the dark energy perturbations. The
		results are displayed in Figure \ref{fig:Pk_Cl}, for a fixed value of $s$. 
		The main physical processes at work can be summarized as follows.
		\begin{itemize}
			\item \emph{Shifted acoustic scale.} The most noticiable effect in the right
				pannel of Figure \ref{fig:Pk_Cl_bgonly} is a shift in the acoustic peaks.
				This is due to the modification of the background expansion history, that
				modifies the angular diameter distance
				\begin{equation}
					d_A = a\int^1_a\frac{\di a'}{a'^2H(a')}\ ,
				\end{equation}
				and hence the acoustic scale, that sets the angular position of the 
				CMB peaks,
				\begin{equation}
					\theta_*=\frac{r_\text{s}(z_*)}{(1+z_*)d_A(z_*)}\ ,
				\end{equation}
				where the comoving sound horizon at decoupling $r_\text{s}(z_*)$ is not
				modified in our case. 
				In particular it will be the phantom behaviour, with $s>0$, that will
				help us to alleviate the Hubble tension.
			\item \emph{Changes in the growth factor.}
				The evolution of the growth factor for matter perturbations is
				governed by, see \eqref{eq:app_GrowthFactorEq},
				\begin{equation}\label{eq:pert_growthFactorEq}
					\ddot{D} + \mathcal{H}\dot{D} - 4\pi a^2 G_\text{eff}\,\rho_m D = 0\ ,
				\end{equation}
				where, in our case, the effective Newton constant is
				\begin{align}\label{eq:pert_Geff}
					\frac{G_\text{eff}}{G} &= 1+\frac{s\mathcal{R}}{3(1+s\mathcal{R})c_A^2}\nonumber\\
						&= 1+ \frac{p_2\mathcal{R}}{\displaystyle 1-p_2\mathcal{R}
							+\frac{2p_2\mathcal{R}}{\phi^2} + \frac{3p_3}{p}\mathcal{X}}\ .
				\end{align}
				Within $\Lambda$CDM, $G_\text{eff}=G$ and the previous equation can be
				solved analitically giving the growing solution
				\begin{equation}\label{eq:pert_growthFactor}
					D(a) \propto H(a)\int^a_0\frac{\di a'}{(a'H(a'))^3}\ .
				\end{equation}
				In the Generalized Proca model, we have two different effects. 
				In the first place, a modification of the expansion history	$H(a)$			
				leads to a modification of the growth factor and it affects the overall 
				amplitude in the matter power spectrum. The reduction in $H(a)$ that can 
				be observed in Figure \ref{fig:H_dA} then leads to an increase in the left 
				pannel of Figure \ref{fig:Pk_Cl_bgonly}, and vice versa.
				
				In the second place, the clustering properties of the dark energy fluid
				are encoded in $G_\text{eff}$. We see that, after taking into account the
				condition $c^2_A>0$, we always have $G_\text{eff}>G$. This effect
				further boosts the amplitude of the matter power spectrum on small scales 
				in Figure \ref{fig:Pk_Cl}.
				
			\item \emph{ISW effect.} 				
				The late-time evolution of the gravitational potentials is affected in two 
				ways. In the first place, if we consider only the modified background, the 
				potentials are still decaying ($\dot{\Phi}<0$) as in $\Lambda$CDM but the 
				overall ISW effect is suppressed.				
				When we include the dark energy perturbations, the late-time evolution
				of the potentials is drastically modified and we can have both growing
				or decaying potentials at late times. As we will comment in 
				Section \ref{sec:ISW}, the case with growing potentials is strongly 
				disfavoured by measurements	of the temperature ISW-galaxy density cross 
				correlation.						
		\end{itemize}				
		
\section{Observations}\label{sec:Observations}
	Once we have discussed the modifications to the background and perturbations, we will
	constrain the parameters of the GP model using cosmological observations and comparing
	with the $\Lambda$CDM predictions. 
	We analyze the following models:
	\begin{itemize}
		\item $\boldsymbol{\Lambda}$\textbf{CDM}. We consider a flat $\Lambda$CDM model with
			fixed neutrino parameters $N_\text{eff}=3.046$ and $\sum m_\nu = 0.06$ eV
			and the usual free parameters $\left\{\Omega_bh^2, \Omega_\text{cdm}h^2,
			100\,\theta_\text{s},\log(10^{10}A_s),n_s,\tau_\text{reio}\right\}$. See 
			\cite{Aghanim:2018eyx} for more details on this parameterization.
		\item \textbf{GP (bg only)}. We consider a GP model without perturbations, taking
			only into account the modified expansion history, where the cosmological 
			constant in the $\Lambda$CDM model is substituted by the Proca field acting 
			as dark energy. On top of the $\Lambda$CDM parameters, this model 
			includes an additional free parameter $\left\{s\right\}$, that modifies the dark energy 
			equation of state \eqref{eq:bg_wA}. We choose a prior $s\in [-0.1, 0.9]$.
		\item \textbf{GP (bg+pert)}. In this model we include both the modified background
			and the dark energy perturbations. On top of the parameters of the previous
			model, we have now two more free parameters $\{\log_{10}(\phi_0), p_2\}$ that
			affect the evolution of the perturbations through their effects on the
			dark energy sound speed \eqref{eq:pert_c2A_def}. We choose priors
			$s\in [0, 0.9]$ and $p_2\in [0, 1]$, consistent with the absence of ghosts
			and strong coupling problems. Additionally, we impose the stability condition
			$c^2_A>0$.
	\end{itemize}
	The sets of observations that we take into account are:
	\begin{itemize}
		\item \textbf{CMB}. We use the full temperature and polarization data (high-$\ell$ TTTEEE 
			and low-$\ell$ TT and EE) from the latest \emph{Planck} 2018 release 
			\cite{Aghanim:2019ame}. We also include all the nuisance parameters.
		\item \textbf{SNe}. We use the Pantheon sample \cite{Scolnic:2017caz}.
		\item \textbf{BAO}. We consider data from BOSS  DR12 \cite{Alam:2016hwk} 
			and WiggleZ \cite{Kazin:2014qga}.
		\item \textbf{HST}. We use the measurement $H_0=74.03\pm 1.42$ 
			from the SH0ES collaboration \cite{Riess:2019cxk}.
	\end{itemize}		
	
	\subsection{Observational constraints}	    
\begin{figure}[ht]
			\centering
			\includegraphics[scale=0.55]{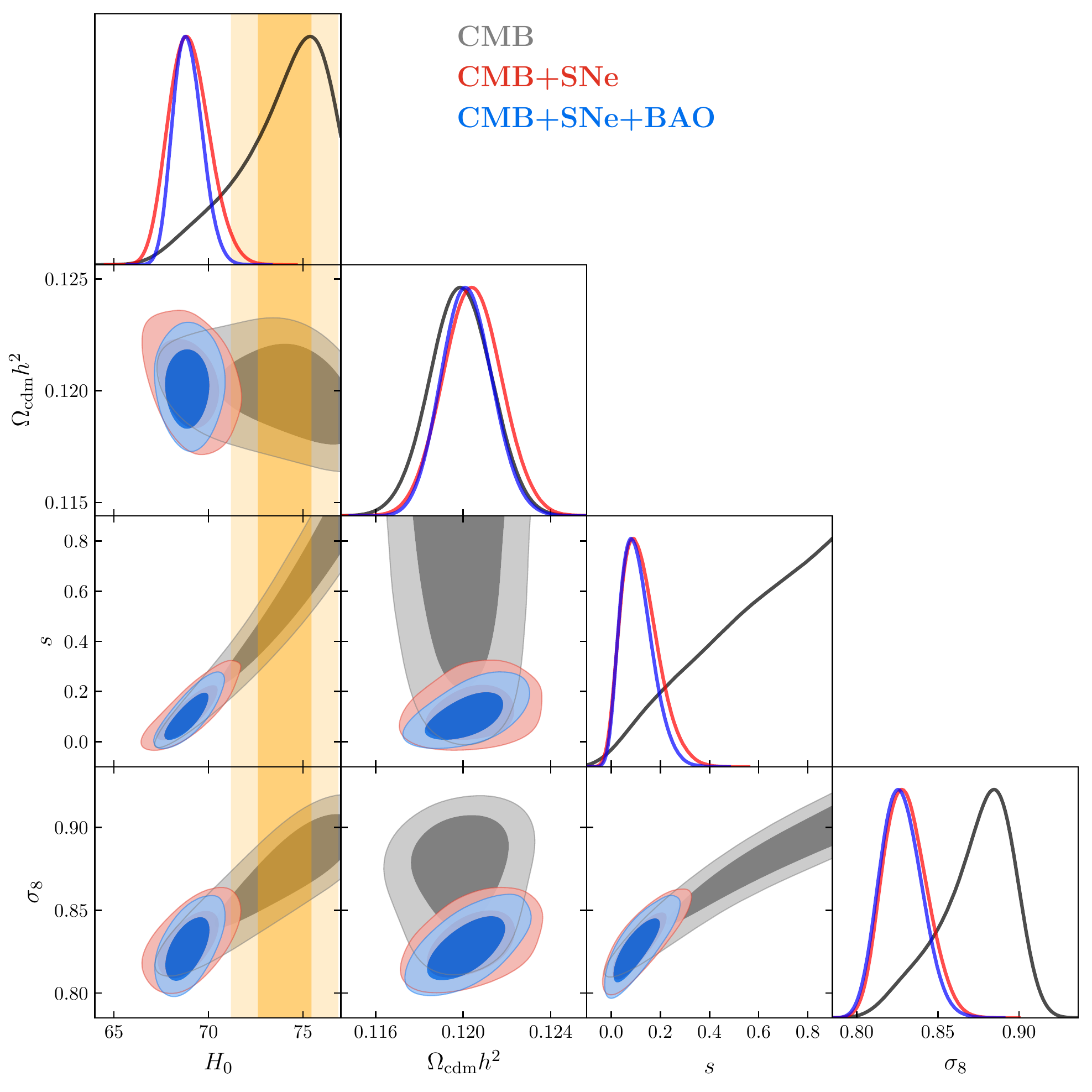}
			\caption{68\% and 95\% CL contours for the GP (bg only) model. The vertical 
			bands are the 68\% and 95\% CL limits from the local distance measurement of 
			the Hubble constant \cite{Riess:2019cxk}.}
			\label{fig:contour_Bg}
		\end{figure}

\begin{figure}[ht]
			\centering
			\includegraphics[scale=0.45]{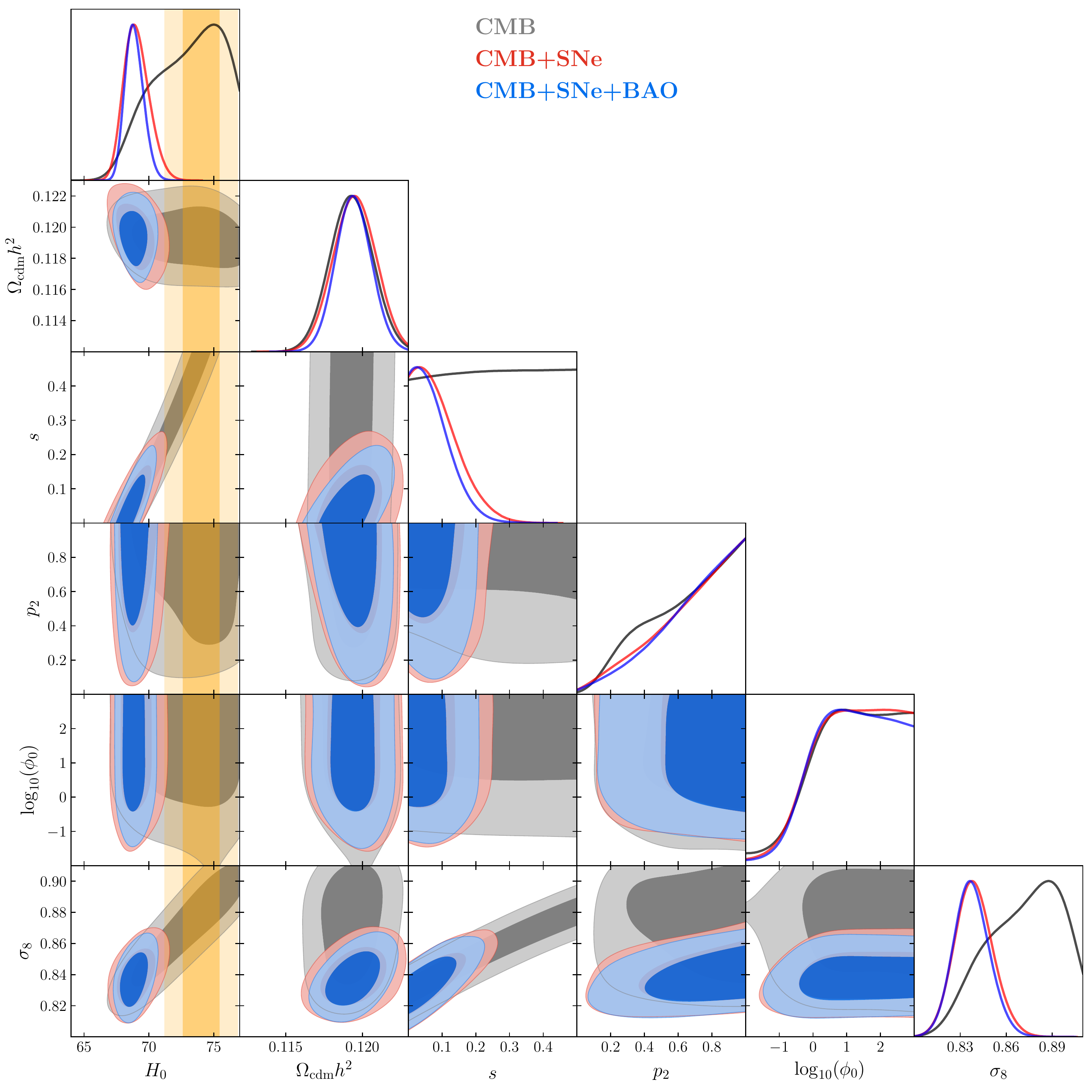}
			\caption{68\% and 95\% CL contours for the GP (bg+pert) model. The vertical 
			bands are the 68\% and 95\% CL limits from the local distance measurement of 
			the Hubble constant \cite{Riess:2019cxk}.}
			\label{fig:contour_Full}
		\end{figure}

		We perform four different runs for each model, combining CMB data with SNe, BAO 
		and the direct $H_0$ measurement. The main results of the fit are collected in 
		Table \ref{tab:cosmo_params}. Figures \ref{fig:contour_Bg} and \ref{fig:contour_Full}
		show the marginalized contours for the two GP models, including the additional 
		GP parameters and the most relevant cosmological variables. We also use the Akaike
		Information Criterion (AIC) \cite{akaike1974new, Liddle:2007fy} to quantify the 
		improvement of GP over the $\Lambda$CDM fit
		\begin{equation}
			\text{AIC} = 2 N_\text{param} + \chi^2\ ,
		\end{equation}
		where $N_\text{param}$ is the number of free parameters of the model and
		$\chi^2=-2\log\mathcal{L}_\text{max}$ is computed from the maximum of the likelihood.
		This information criterion allows us to compare two competing models, penalizing 
		models with more free parameters. For two different models $A$ and $B$, 
		$\text{AIC}_A-\text{AIC}_B>5$ is widely regarded as a strong preference for 
		model $B$ \cite{Liddle:2007fy}.
						
		In the Figures \ref{fig:contour_Bg} and \ref{fig:contour_Full}, we only show the results \emph{without} HST data. As we can see, 
		it is crucial to include SNe data, that constrains the equation of state of dark energy, in order to break
		the large degeneracy between $s$ and $H_0$. Another way to break this degeneracy
		is to use the late-time measurements of $H_0$ which would pull both $s$ and $H_0$ 
		toward higher values.
		
		Once we include the perturbations in the GP model, we find that the parameters 
		$p_2$ and $\phi_0$ cannot be very well constrained. These two parameters affect
		the evolution of the perturbations through their effect on the sound speed and
		$G_\text{eff}$, see \eqref{eq:pert_Geff}. From \eqref{eq:pert_Geff} we can see
		that in the limit $\phi_0\to 0$ we recover the $\Lambda$CDM behaviour, with
		$G_\text{eff}\to G$. However, the results in Figure \ref{fig:contour_Full}
		show that \emph{large} values of $\phi_0$ are actually favoured. We have checked 
		that this is due to the fact that, in this region of parameter space, the 
		suppression of the CMB plateau leads to a better fit to the low-$\ell$ TT data.
		As can be checked in Table \ref{tab:cosmo_params}, this model improves the 
		$\Lambda$CDM fit even without including the HST observation. However, also in
		this region of parameter space, the late-time growth of the metric potentials may
		result in a strong disagreement with the observations of temperature ISW and 
		galaxy density cross-correlations, not included in our analysis. We will discuss 
		this matter further in Section \ref{sec:ISW}.  	
											
	    \begin{table}
	    	\centering
	    	{\renewcommand{\arraystretch}{1.3}
		    \begin{tabular}{l|ccc|ccc}
				&\multicolumn{3}{c|}{CMB+SNe+BAO}
				           &\multicolumn{3}{c}{CMB+SNe+BAO+HST}\\\hline
				                       & $\Lambda$CDM & GP (bg)     & GP (bg+pert) 
				                       & $\Lambda$CDM & GP (bg)     & GP (bg+pert) \rule{0pt}{2.5ex} \\\hline
				$100\, \Omega_b h^2$     & $2.244\pm 0.014$ & $2.237\pm 0.014$ & $2.246\pm 0.015$
						               & $2.255\pm 0.013$ & $2.239\pm 0.015$ & $2.248\pm 0.015$\\
				$\Omega_\text{cdm}h^2$ & $0.1191\pm 0.001$    & $0.1202\pm 0.0012$   & $0.1194\pm 0.0012$
						               & $0.1179\pm 0.00096$  & $0.1202\pm 0.0012$   & $0.1195\pm 0.0013$\\
     		    $100\,\theta_\text{s}$ & $1.042\pm 0.00029$   & $1.042\pm 0.00029$   & $1.042\pm 0.00029$
						               & $1.042\pm 0.00029$   & $1.042\pm 0.0003$    & $1.042\pm 0.0003$\\
				$\log(10^{10}A_s)$     & $3.047\pm 0.017$     & $3.046\pm 0.016$ & $3.040\pm 0.016$
				                       & $3.048\pm 0.017$     & $3.045\pm 0.017$ & $3.040\pm 0.016$\\
				$n_s$                  & $0.9679\pm 0.0039$   & $0.9653\pm 0.0041$ & $0.9671\pm 0.0042$
				                       & $0.9709\pm 0.0038$   & $0.9653\pm 0.0042$ & $0.967\pm 0.0044$\\
				$\tau_\text{reio}$     & $0.05617_{-0.0084}^{+0.0076}$   & $0.05437_{-0.0083}^{+0.0076}$ & $0.05267\pm 0.0078$
				                       & $0.05805_{-0.0087}^{+0.0074}$   & $0.05406_{-0.0082}^{+0.0077}$ & $0.05258_{-0.0079}^{+0.0076}$\\\hline
				$s$ 		           & -- & $0.1049_{-0.083}^{+0.043}$ & $0.07588_{-0.076}^{+0.019}$
				                       & -- & $0.199_{-0.083}^{+0.065}$  & $0.1613_{-0.081}^{+0.066}$\\
				$p_2$                  & -- & -- & $0.6916_{-0.085}^{+0.31}$
				                       & -- & -- & $0.6887_{-0.09}^{+0.31}$\\
				$\log_{10}(\phi_0)$    & -- & -- & No constraint
				                       & -- & -- & No constraint\\\hline
			    $\chi^2$               & 3804 & 3804 & 3794
							           & 3822 & 3814 & 3804 \\
				$\Delta$AIC 
				\rule{0pt}{2.5ex}      & 0 & -2 & 4
				  	                   & 0 &  6 & 12
	    	\end{tabular}}
	    	\caption{Constraints on the free parameters of the models, considering two
	    	datasets: with and without the local distance measurement of the Hubble constant. 
	    	We also include the minimum of the $\chi^2$ and a comparison of the GP models
	    	to $\Lambda$CDM using the AIC, computed as 
	    	$\Delta\text{AIC}=\text{AIC}_{\Lambda \text{CDM}} - \text{AIC}_{\text{GP}}$. 
	    	The degenaracy in the $p_2$-$\phi_0$ plane is too large to obtain a credible 
	    	confidence interval for $\phi_0$. As could be anticipated, both GP models can
	    	accomodate better the HST measurement of $H_0$ providing a better fit than
	    	$\Lambda$CDM.}
	    	\label{tab:cosmo_params}
	    \end{table}
	    
	\subsection{$H_0$ tension}
		As we have already discussed, the phantom-like equation of state of GP models 
		modifies the angular diameter distance to the last-scattering surface, shifting the CMB
		peaks and favouring a large $H_0$. Other models of dark energy 
		with phantom-like equations of state, like scalar Galileons 
		\cite{Barreira:2014jha, Renk:2017rzu}, present a 
		similar behaviour. When comparing with CMB observations, this effect introduces a 
		large degeneracy between the 
		parameter $s$, that governs the dark energy equation of state, and $H_0$. Adding 
		external data is crucial to break this degeneracy. 
		
\begin{figure}[ht]
			\centering
			\includegraphics[scale=0.75]{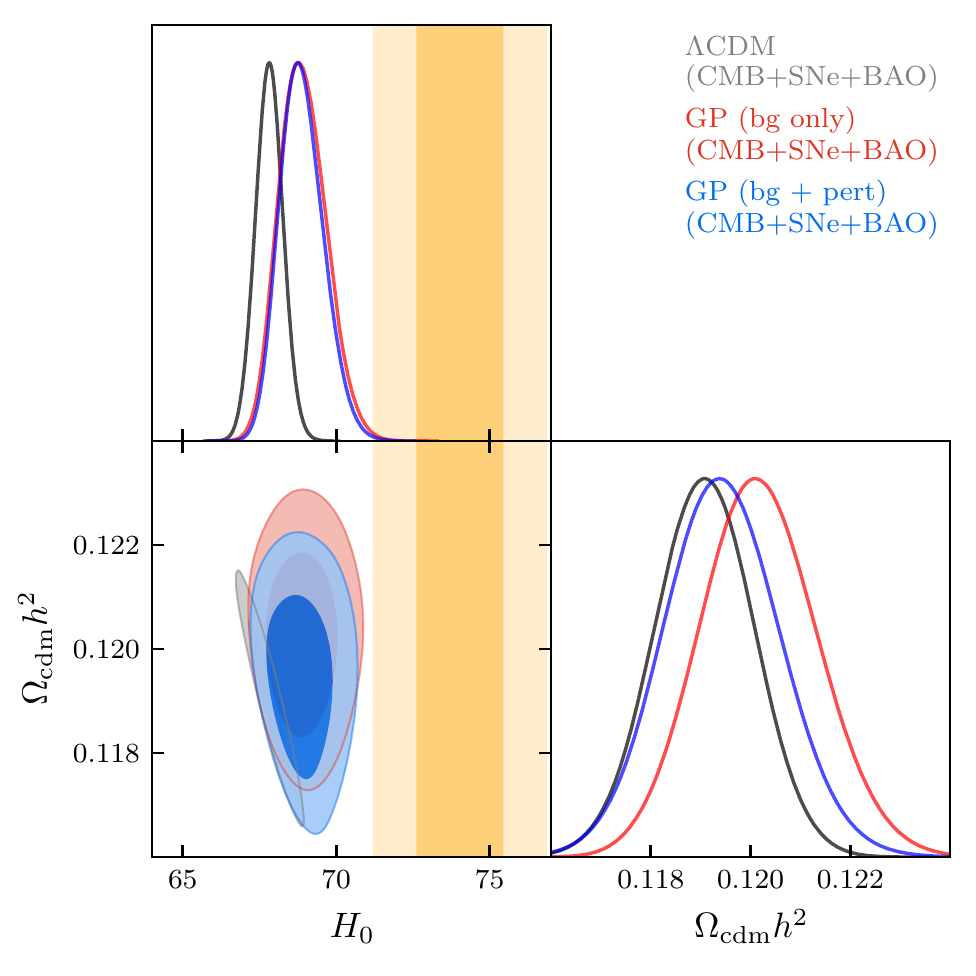}
			\caption{Comparison of the $\Lambda$CDM and GP models
			(with 68\% and 95\% CL). Again, the vertical band
			is the HST measurement of $H_0$ \cite{Riess:2019cxk}. Notice that in this
			plot, the HST has \emph{not} been included in the analysis. Even without
			the HST prior, GP models predict a larger $H_0$. Once we include HST, the 
			fit is pulled towards even larger values of $H_0$, close to the local value.}
			\label{fig:comparison_CMB+SNe+BAO}
		\end{figure}

		One possibility is to use a direct measurement of the Hubble constant, like the 
		local value from the SH0ES collaboration \cite{Riess:2019cxk}. This was the 
		approach taken in \cite{DeFelice:2020sdq}. However, in order
		to address the so-called $H_0$ tension without imposing this prior, we have chosen 
		to break this degeneracy combining CMB with SNe and BAO data. When we include the 
		HST measurement as well, $H_0$ is pulled toward larger values.		
		The most relevant results for the Hubble tension are collected in Table 
		\ref{tab:LSS_table} and in Figure \ref{fig:comparison_CMB+SNe+BAO}.
		
		It is important to notice that GP models naturally prefer higher values
		of $H_0$, even without using the HST observation. In this way, GP can fit
		CMB, BAO and SNe data as successfully as $\Lambda$CDM (or even better)
		while reducing the Hubble tension by 1$\sigma$. In addition, if we include 
		HST data, the tension is reduced to $2.4\sigma$.
		
		\begin{table}
	    	\centering
	    	{\renewcommand{\arraystretch}{1.3}
		    \begin{tabular}{l|ccc|ccc}
				&\multicolumn{3}{c|}{CMB+SNe+BAO}
				           &\multicolumn{3}{c}{CMB+SNe+BAO+HST}\\\hline
				           & $\Lambda$CDM & GP (bg) & GP (bg+pert) 
				           & $\Lambda$CDM & GP (bg) & GP (bg+pert) \rule{0pt}{2.5ex} \\\hline
				$\Omega_m$ & $0.3077\pm 0.006$ & $0.3003_{-0.0071}^{+0.0074}$ & $0.2991\pm 0.007$
						   & $0.3005_{-0.0057}^{+0.0054}$ & $0.2894_{-0.0067}^{+0.0062}$ & $0.2889_{-0.0066}^{+0.0063}$ \\
				$\sigma_8$ & $0.8088_{-0.0076}^{+0.0072}$ & $0.8273\pm 0.013$ & $0.8374\pm 0.012$
						   & $0.8059_{-0.0077}^{+0.0072}$ & $0.8397\pm 0.013$ & $0.8477\pm 0.013$ \\
     		    $S_8$      & $0.819\pm 0.013$  & $0.8276\pm 0.014$ & $0.8362\pm 0.014$
						   & $0.8066\pm 0.013$ & $0.8245\pm 0.013$ & $0.8319\pm 0.014$\\
				$H_0$      & $67.83\pm 0.45$   & $68.91_{-0.86}^{+0.69}$ & $68.88_{-0.8}^{+0.62}$
				           & $68.38\pm 0.43$   & $70.21\pm 0.76$ & $70.1\pm 0.76$ \\\hline
				$H_0$ tension & $4.2\sigma$ & $3.2\sigma$ & $3.2\sigma$
				              & $3.8\sigma$ & $2.4\sigma$ & $2.4\sigma$
	    	\end{tabular}}
	    	\caption{Constraints on some derived parameters, considering two
	    	datasets: with and without the local distance measurement of the Hubble 
	    	constant. The Hubble tension is evaluated approximating the 1d posterior on 
	    	$H_0$ as Gaussian and comparing with the local value $H_0=74.03\pm 1.42$ 
			from the SH0ES collaboration \cite{Riess:2019cxk}.}
	    	\label{tab:LSS_table}
	    \end{table}		
		
	\subsection{Complementary observations: $\sigma_8$ and ISW effect}\label{sec:ISW}
		The novel clustering properties of the dark energy fluid can produce, for some
		combinations of parameters, a late-time growth (deepening) of the gravitational
		potentials. This leads to an increased clustering amplitude, i.e. larger
		$\sigma_8$, and to an ISW effect with the opposite sign, as compared with 
		$\Lambda$CDM. These are two common side effects of late-time solutions to the 
		Hubble tension based on dark energy. While relieving the $H_0$ tension, we can 
		spoil the fit to other observations, e.g. increasing the $\sigma_8$	tension.
		
		\begin{figure}[ht]
			\centering
			\includegraphics[scale=0.45]{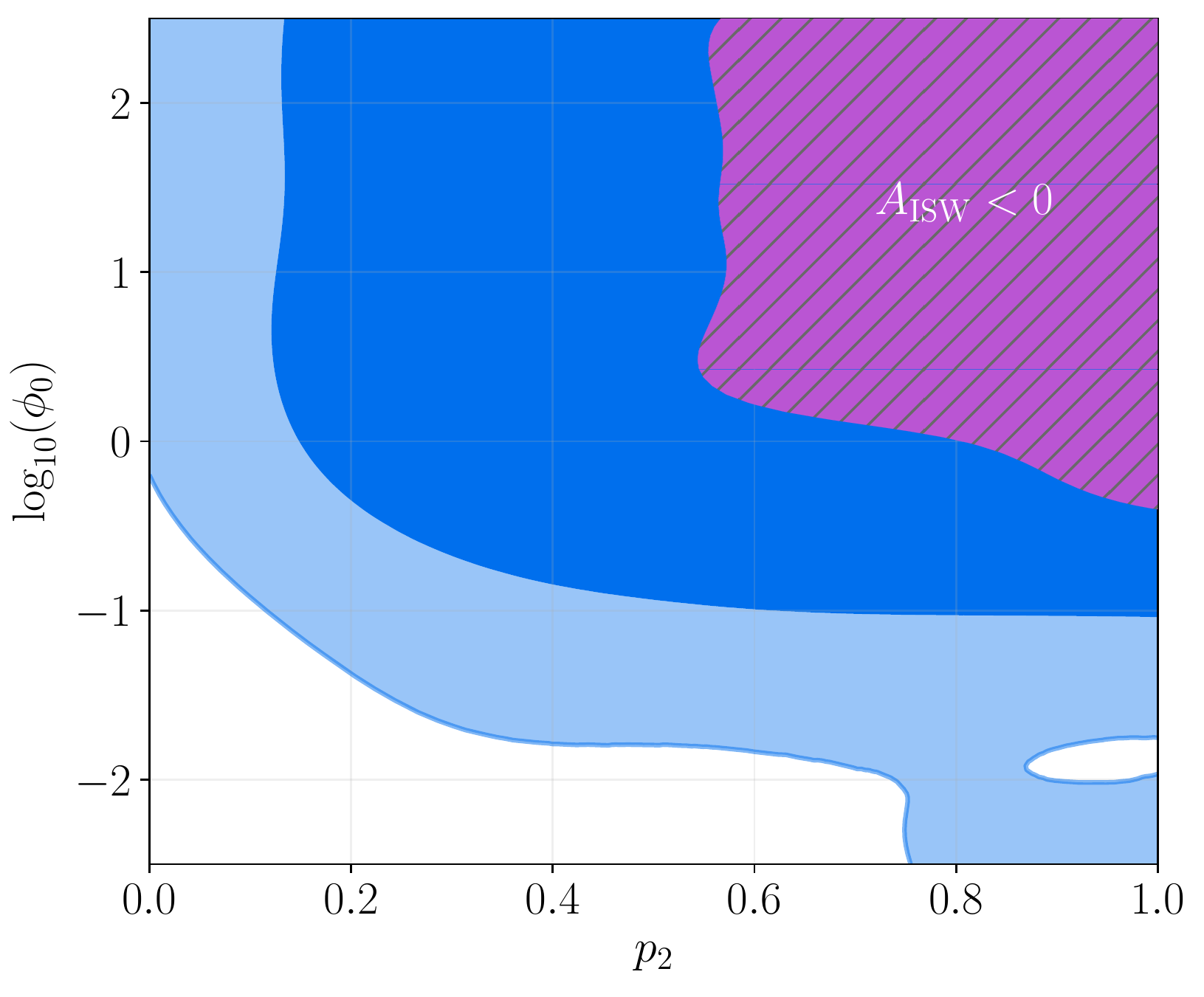}
			\caption{95\% and 99\% CL contour on the GP parameters $\phi_0$-$p_2$. These 
			two parameters only affect the evolution of the perturbations, modifying the
			sound speed of dark energy. The region with $A_\text{ISW}<0$ is
			most likely excluded by measurements of the temperature ISW-galaxy density
			cross-correlation.}
			\label{fig:contour_ISW}
		\end{figure}
		
		\begin{figure}[ht]
			\centering
			\includegraphics[scale=0.65]{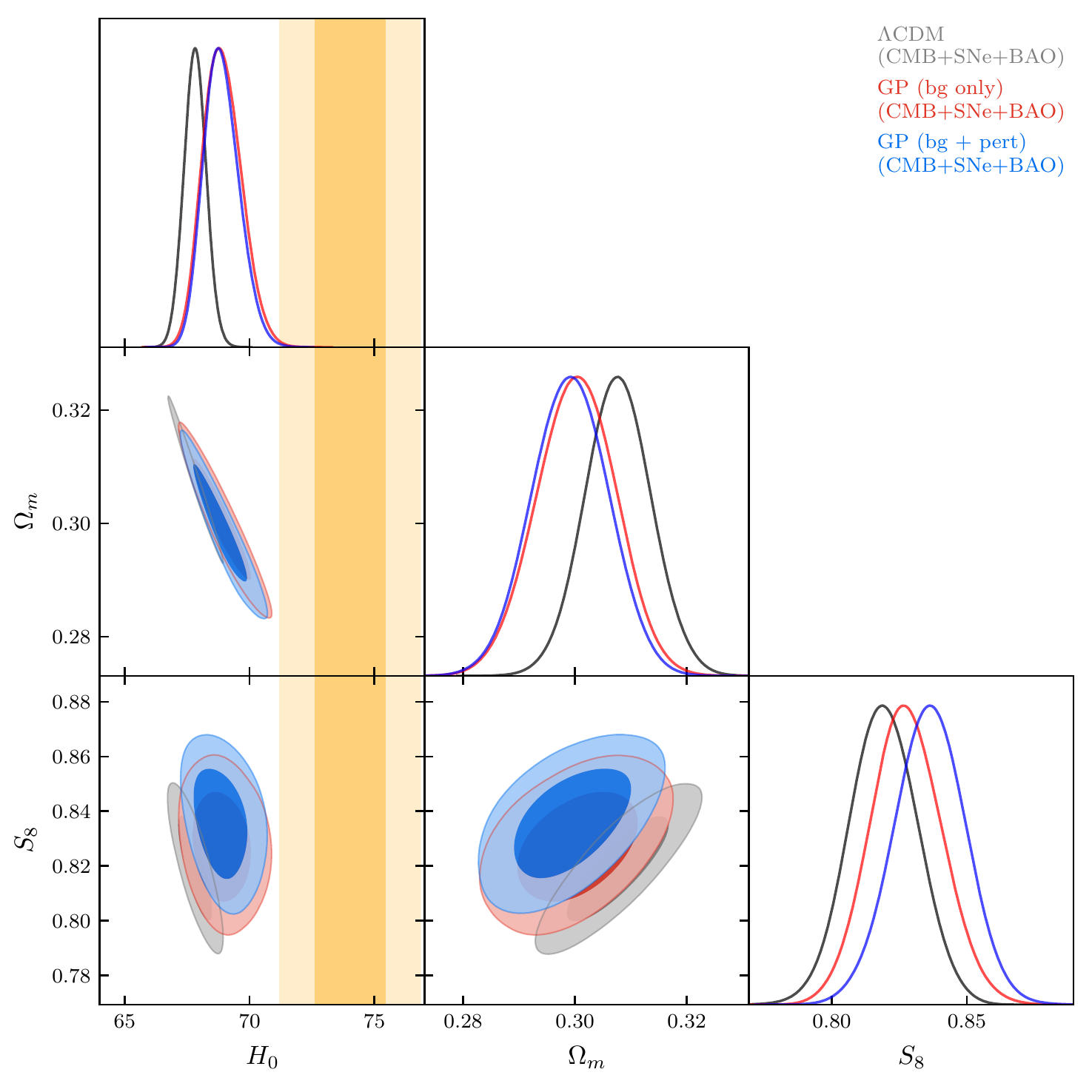}
			\caption{Comparison of the $\Lambda$CDM and GP models (with 68\% and 95\% CL). 
			We use the variable $S_8\equiv \sigma_8\sqrt{\Omega_m/0.3}$, that is very
			precisely constrained by weak lensing surveys. On top of predicting a higher
			$H_0$ value, GP models also favour lower values of $\Omega_m$.}
			\label{fig:S8_CMB+SNe+BAO}
		\end{figure}
		
		The consistency of the $\sigma_8$ value inferred from the CMB and the one
		measured by weak lensing and galaxy clustering surveys has been a matter of
		debate for years, and is usually known as the $\sigma_8$ tension.
		On the one hand, the latest \emph{Planck} value (including CMB lensing and BAO data) for the 
		parameter $S_8\equiv \sigma_8\sqrt{\Omega_m/0.3}$ is 
		$S_8=0.825\pm 0.011\ (68\%\ \text{CL})$ \cite{Aghanim:2018eyx}. On the other hand, 
		the latest measurements 
		by DES Y1 ($S_8=0.773^{+0.026}_{-0.020}$) \cite{Abbott:2017wau} 
		and KiDS-1000 ($S_8=0.766^{+0.020}_{-0.014}$) \cite{heymans2020kids}
		are consistently lower (see Fig. 5 in \cite{heymans2020kids} for a full list of 
		recent measurements).
		
		Even though the different collaborations do not seem to agree about the consistency
		of the results, i.e. about the existence of a `tension', a general lesson to be 
		learnt is that any $S_8$ value significantly higher than the \emph{Planck}
		measurement should be strongly disfavoured. It would then seem worrying that in
		the GP model we obtain slighty larger values for $S_8$, see Table 
		\ref{tab:LSS_table}, but there are at least two caveats. First, the constraints
		from galaxy surveys depend on the underlying cosmological model. For instance, the DES 
		collaboration also reports the measurement $S_8=0.782^{+0.036}_{-0.024}$ in a dark 
		energy model with a constant equation of state ($w$CDM). To perform a fair comparison
		with the GP model, one would need to take into account the modified background.
		The second caveat is that, to adress the consistency of measurements, it is
		important to analyze the multidimensional posterior. Besides decreasing $S_8$, 
		another generic feature of
		weak lensing surveys is to predict lower values for $\Omega_m$, thus displacing
		the minimum in the $\Omega_m$-$S_8$ plane. While the GP model increases slightly 
		the tension in one direction (increasing $S_8$) it also predicts lower values 
		for $\Omega_m$, reducing the tension in that direction. All these effects 
		should be taken into account when comparing with LSS clustering data.
				
		The temperature ISW effect depends on the (integrated) time evolution of the 
		gravitational potentials. The low-$\ell$ CMB plateau is affected by modifications
		to the ISW term, but it is largely insensitive to its sign. This sign carries 
		important information, e.g. it can reveal whether the potentials are growing
		or decaying, and it can be recovered from the cross-correlation of CMB temperature
		data and galaxy density observations 
		\cite{Bean:2003fb, Hu:2004yd, Renk:2017rzu, Ballardini:2018cho}. 
		Different observations are available and have been fruitfully applied to test
		dark energy \cite{Stolzner:2017ged, Nakamura:2018oyy, Giacomello:2018jfi}. In 
		particular, ISW observations have already been used to constrain our GP model
		\cite{Nakamura:2018oyy} but without using the full CMB data. The results of 
		\cite{Nakamura:2018oyy} show that small values of $\phi_0$ are preferred by these 
		observations. 
		
		A precise comparison with ISW data would require a dedicated analysis. In this
		work, we only estimate the sign of the ISW effect,
		by adding all the cross-correlation multipoles, 
		$A_\text{ISW}\equiv \sum_\ell C^{Tg}_\ell$. Combinations of	parameters that yield
		a negative cross-correlation ($A_\text{ISW}<0$) must be strongly disfavoured. In 
		Figure \ref{fig:contour_ISW} we show the 2$\sigma$ and 3$\sigma$ contours of the two
		parameters related to the clustering properties of the model, $\phi_0$ and $p_2$.
		In agreement with \cite{Nakamura:2018oyy}, we observe that a large region of parameter 
		space, with large $\phi_0$ values, is ruled out by ISW observations. The 
		combination of the full	CMB data, that favours large $\phi_0$ values as we show in 
		this work, and temperature ISW-galaxy density cross-correlation, that favours 
		small $\phi_0$ values as shown in \cite{Nakamura:2018oyy}, could then be used to 
		break the degeneracy and tightly constrain $\phi_0$.

\section{Summary and conclusions}\label{sec:Conclusions}
	In this work, we have studied in detail the phenomenology of a simple Generalized
	Proca model for dark energy. After comparing with CMB, BAO and SNe observations 
	(without imposing a $H_0$ prior) we	have showed that it can relieve the Hubble tension
	by about $1\sigma$.
	
	This class of models can reduce the Hubble tension mainly because of
	the phantom-like behaviour of dark energy, in a similar way to scalar Galileon models.
	The modified expansion history reduces the angular diameter distance to the 
	last-scattering surface, shifting the CMB peaks and introducing a high degeneracy 
	with $H_0$. In this scenario, if we impose a prior in $H_0$, like in \cite{DeFelice:2020sdq}, 
	a high value is trivially preferred. However, in our case, instead of using the 
	local $H_0$ measurements,
	we have broken this degeneracy in the CMB data using SNe data, that can constrain the
	equation of state of dark energy. We have shown that, even without the local $H_0$
	prior, the GP model naturally favours higher values of $H_0$, easing the Hubble tension.
	
	The modified background is the main responsible for relieving the Hubble tension, but
	we have also studied the perturbations in the model. Both the novel clustering properties 
	of dark energy and the modified background produce a late-time evolution of the metric
	potentials that modify the ISW effect. This produces a modification on the CMB spectrum
	at low $\ell$, i.e. the SW plateau. In particular, once we include the perturbations,
	the supression of the SW plateau can lead to a better fit to CMB data. 

	Finally, we checked the consistency of our results with complementary observations
	from LSS. A well-known side effect of dark-energy solutions to the Hubble tension
	is a potential disagreement with LSS observations. The GP model predicts a
	slightly larger value for $\sigma_8$, while also reducing the value of $\Omega_m$
	so the tension on the $\sigma_8$-$\Omega_m$ plane would have to be properly studied, 
	and can also potentially produce an ISW-galaxy density cross-correlation with the wrong
	sign. Some of these aspects have been previously studied in \cite{Nakamura:2018oyy}, 
	where the authors constrained the GP model using ISW measurements, but only using 
	reduced CMB information. It was shown that the GP model is compatible with ISW 
	observations in some region of parameter space. 
	
	Further study in this direction would involve extending our analysis to include 
	clustering data and ISW measurements. The compromise between achieving a better 
	fit to CMB low-$\ell$, as shown in this work, while simultaneously agreeing with ISW 
	observations, as in \cite{Nakamura:2018oyy}, would then allow us to confidently
	constrain all the parameters of this model.

\begin{acknowledgments}
LH is supported by funding from the European Research Council (ERC) under the European Unions Horizon 2020 research and innovation programme grant agreement No 801781 and by the Swiss National Science Foundation grant 179740. 
\end{acknowledgments}

\appendix
\section{General perturbations}\label{app:gen_pert}
	In this appendix we present the full equations of motion and the energy-momentum tensor, 
	for a generic metric and field perturbation and before applying the equations of motion
	at the background level.
	We start with a general perturbed FLRW metric, that can be written as 
	\begin{equation}
		\di s^2 = a^2(\tau)\Big(-(1-A)\di\tau^2 + 2B_i\di\tau\di x^i 
			+ (\delta_{ij} + H_{ij})\di x^i\di x^j\Big)\ ,
	\end{equation}
	and we parameterize the perturbations to the Proca field as
	\begin{subequations}
	\begin{align}
		A^0 &= \frac{1}{a}(\phi + \delta\phi)\ ,\\
		A^i &= a^{-2}\delta^{ij}(\partial_j\chi_V + E_j)\ ,
	\end{align}
	\end{subequations}
	where $\delta\phi$ and $\chi_V$ are scalar perturbations and $E_i$ is the transverse 
	part of the vector field, i.e. a vector perturbation. \\
	
	\paragraph*{Equations of motion.}
		\begin{align}
			\mathcal{E}_0 &= 
			\alpha\frac{k^2}{a^2}\left\{a\delta\phi + \dot{\chi}_V - a\frac{\ii \hat{k}^i}{k}B_i\dot{\phi}
				-a\phi\left[A+\frac{\ii\hat{k}^i}{k}\left(\dot{B}_i+\mathcal{H}B_i
				\right)\right]\right\}\nonumber\\
			&\quad + \left(\delta\phi-\frac{1}{2}\phi A\right)\bigg\{a G_{2,X} 
				+aG_{2,XX}\phi^2 + 3\mathcal{H}G_{3,XX}\phi^3+6\mathcal{H}G_{3,X}\phi\bigg\}\nonumber\\
			&\quad -\frac{1}{2}aAG_{2,X}\phi 
				- a^{-1}G_{3,X}\phi\left(k^2\chi_V - \frac{1}{2}a\phi\dot{H}\indices{^i_i}\right)=0\ ,
		\end{align}
		\begin{align}
			\mathcal{E}_i &=
			-\frac{\alpha}{a^2}\bigg\{
				a\ii k_i\left(\delta\dot{\phi}-\phi\dot{A}+\mathcal{H}\delta\phi
					-\mathcal{H}\phi A-\dot{\phi}A\right)
				+a\phi\ddot{B}_i+2a\dot{B}_i(\dot{\phi}+\mathcal{H}\phi)-a\phi k_ik^jB_j\nonumber\\
				&\qquad\qquad +aB_i\left(\ddot{\phi}+2\mathcal{H}\dot{\phi}
					\left(\dot{\mathcal{H}}+\mathcal{H}^2+k^2\right)\phi\right)
				+\ii k_i\ddot{\chi}_V + \ddot{E}_i + k^2E_i\bigg\}\nonumber\\
			&\quad -G_{2,X}\left(\ii k_i\chi_V + E_i + a\phi B_i\right)\nonumber\\
			&\quad-G_{3,X}\bigg\{\ii k_i\phi\left(\delta\phi - \frac{1}{2}\phi A\right)
				+a^{-1}\left(\dot{\phi}+3\mathcal{H}\phi\right)
				\left(\ii k_i\chi_V+E_i+a\phi B_i\right)\bigg\}=0\ .
		\end{align}		
	
	\paragraph*{Energy-momentum tensor.}
		\begin{align}
			T\indices{^0_0} &= -\frac{\phi}{a}\left(\delta\phi - \frac{1}{2}\phi A\right)
				\left(aG_{2,X}+aG_{2,XX}\phi^2 + 9\mathcal{H}G_{3,X}\phi
					+3\mathcal{H}G_{3,XX}\phi^3\right)\nonumber\\
				&\quad - \frac{3}{2}\frac{\mathcal{H}}{a}AG_{3,X}\phi^3
					+ \frac{\phi^2}{a^2}G_{3,X}\left(k^2\chi_V 
					- \frac{1}{2}a\phi\dot{H}\indices{^i_i}\right)\ ,
		\end{align}
		\begin{align}
			T\indices{^0_i} &= \frac{\phi}{a^2}\left(a G_{2,X} + 3\mathcal{H}\phi G_{3,X}\right)
				\big(\ii k_i\chi_V + E_i + a\phi B_i\big)\nonumber\\
			&\quad +\frac{\ii k_i}{a}G_{3,X}\phi^2\left(\delta\phi-\frac{1}{2}\phi A\right)\ ,
		\end{align}
		\begin{align}
			T\indices{^i_j} &= -\frac{\phi}{a}\delta\indices{^i_j}\bigg\{\phi G_{3,X}
				\left(\delta\dot{\phi}-\frac{1}{2}\phi \dot{A}\right) 
				-\left(\delta\phi -\frac{1}{2}\phi A\right)
				\left(aG_{2,X}-2\dot{\phi}G_{3,X}-\dot{\phi}G_{3,XX}\phi^2\right)\bigg\}\ .
		\end{align}
		
	It is important to notice that once we apply the equations of motion at the background
	level, the vector component of $T\indices{^0_i}$ vanishes. This implies that the
	vector field does not source vorticity in the Einstein equations and thus it is
	purely decaying and negligible as in $\Lambda$CDM (in agreement with \cite{DeFelice:2016uil}). Once we apply the equations of 
	motion at the background level, \eqref{eq:bg_eom0} and \eqref{eq:bg_constraint2}, the
	results of the main text can be recovered setting
	\begin{align}
		A &= -2\Psi\ ,\\
		B_i &= \ii k_i B\ ,\\
		H_{ij} &= -2\Phi\delta_{ij} -2k_ik_j E\ .
	\end{align}
	
\section{Super- and sub-Hubble analysis}
	In this appendix we will use $N=\log(a)$ as time variable, and denote $'\equiv \partial_N$.
	The energy-momentum tensor conservation for a generic fluid leads to the continuity
	and Euler equations
	\begin{subequations}\label{eq:app_FluidEqs}
	\begin{align}
		\Delta' &= 3w\Delta - \frac{k}{\mathcal{H}}\left\{1+\frac{3\mathcal{H}}{k}
			\left(1-\frac{\mathcal{H}'}{\mathcal{H}}\right)\right\}u - 4\sigma 
			+ (1+w)\left(\Phi'+\Psi\right)\ ,\\
		u' &= -u - 3(c_\text{s}^2-w)u -+ \frac{kc_\text{s}^2}{\mathcal{H}}\Delta - \frac{4k}{3\mathcal{H}}\sigma
			+ \frac{k}{\mathcal{H}}(1+w)\Psi\ ,
	\end{align}
	\end{subequations}
	expressed in terms of the gauge-invariant density perturbation $\Delta$ and the 
	velocity perturbation $u$, that are related to the usual energy density and velocity
	divergence as
	\begin{equation}
		\Delta\equiv \delta + \frac{3\mathcal{H}(1+w)\theta}{k^2}\ ,\qquad
		u \equiv \frac{1}{k}(1+w)\theta\ .
	\end{equation}
	The Einstein equations can be written as
	\begin{subequations}\label{eq:app_EinsEqs}
	\begin{align}
		\Phi &= -\frac{3\mathcal{H}^2}{2k^2}\left(\mathcal{R}\mathcal{Z}
			+\tilde{\mathcal{R}}\Delta\right)\ ,\\
		\Phi-\Psi &= \frac{6\mathcal{H}^2}{k^2}\mathcal{R}\sigma\ ,\\
		\Phi'+\Psi &= \frac{3\mathcal{H}}{2k(1+s\mathcal{R})}\left(\frac{k}{3\mathcal{H}}\mathcal{R}\mathcal{Q}
			+\tilde{\mathcal{R}}u\right)\ .
	\end{align}
	\end{subequations}
	For simplicity, we will restrict ourselves to perfect fluids, $\sigma=0$, and 
	$c_\text{s}^2=w=\text{const.}$. Using the Einstein equations \eqref{eq:app_EinsEqs}, 
	we can combine the equations for the fluid \eqref{eq:app_FluidEqs} and for the Proca 
	field \eqref{eq:pert_eoms} into two coupled second-order differential 
	equations for the gauge-invariant density perturbations $\mathcal{Z}$ and $\Delta$.
	In order to grasp the behaviour of the system, we will study both the super-Hubble
	($k\ll \mathcal{H}$) and sub-Hubble ($k\gg \mathcal{H}$) limits.
	
	\subsection{Super-Hubble limit}
		In the super-Hubble limit, the evolution of $\mathcal{Z}$ decouples from $\Delta$, 
		i.e. $\Delta$ does not appear in the equation for $\mathcal{Z}'$. However, as we
		will see later, the super-Hubble behaviour of $\mathcal{Z}$ is not particularly
		relevant since there is an attractor solution in the sub-Hubble regime. The 
		full expressions in this regime are not very illuminating, so we will focus on 
		two limiting cases. First, if the Proca field is subdominant ($\mathcal{R}\to 0$)
		we have
		\begin{align}
			\mathcal{Z}'' &= -5\left(1-\frac{3(1+w)(1-2s)}{10}\right)\mathcal{Z}'
				- 6\left\{1+(1+w)\left(s + \frac{3}{4}\big(c^2_A-1-s(1+w)\big)\right)\right\}\mathcal{Z}
				+ \Od{\frac{k}{\mathcal{H}}}\ ,\\
			\Delta'' &= -\frac{9w-1}{2}\Delta' - \frac{3}{2}(w-1)(1+3w)\Delta
				+ \Od{\frac{k}{\mathcal{H}}}\ .
		\end{align}
		In the opposite limit, on the de Sitter attractor $\mathcal{R}\to 1$, we have
		\begin{align}
			\mathcal{Z}'' &= -(15 +9w)\mathcal{Z}' - (8+3w)\mathcal{Z}
				+ \Od{\frac{k}{\mathcal{H}}}\ ,\\
			\Delta'' &= (3w-2)\Delta' + 6w\Delta
				- 3(1+w)\mathcal{Z}' + \frac{3}{2}(1+w)\left(3c^2_A-5\right)\mathcal{Z} 
				+ \Od{\frac{k}{\mathcal{H}}}\ .
		\end{align}
		
	\subsection{Sub-Hubble limit}
		To lowest order in the sub-Hubble limit we have
		\begin{align}
			\mathcal{Z}'' &= -\frac{k^2}{\mathcal{H}^2}\left\{c^2_A\mathcal{Z}
				-\frac{s(1+3w)}{3(1+s\mathcal{R})}\tilde{\mathcal{R}}\Delta\right\}
				+ \Od{1}\ ,\\
			\Delta'' &= -\left(1-3w+\frac{\mathcal{H}'}{\mathcal{H}}\right)\Delta'
				+ \left(1+3w-(1-3w)\frac{\mathcal{H}'}{\mathcal{H}}\right)\Delta
				- \frac{k^2}{\mathcal{H}^2}w\Delta\nonumber\\
			&\quad + \frac{3}{2}(1+w)(1+3c^2_A)\mathcal{Z} 
				+ \Od{\frac{\mathcal{H}}{k}}\ .
		\end{align}
		In this regime we can approximate
		\begin{equation}
			\mathcal{R}\mathcal{Z}\simeq (1+3w)\mathcal{F}\tilde{\mathcal{R}}\Delta\ ,\qquad
			\mathcal{F}\equiv \frac{s\mathcal{R}}{3(1+s\mathcal{R})c^2_A}\ .
		\end{equation}
		Substituting into the equation for $\Delta$
		\begin{equation}
			\Delta'' = -\left(1-3w+\frac{\mathcal{H}'}{\mathcal{H}}\right)\Delta'
				+\frac{3}{2}\tilde{\mathcal{R}}(1+3w)(1+w)(1+\mathcal{F})\Delta
				+6\frac{\mathcal{H}'}{\mathcal{H}}w\Delta -\frac{k^2}{\mathcal{H}^2}w\Delta\ .
		\end{equation}
		In particular, in the late Universe $w=0$ and $\Delta$ is the total matter density perturbation.
		In this case, we obtain the usual expression for the matter growth function
		\begin{equation}\label{eq:app_GrowthFactorEq}
			\Delta'' = -\left(1+\frac{\mathcal{H}'}{\mathcal{H}}\right)\Delta'
				+ \frac{4\pi G_\text{eff}\, a^2}{\mathcal{H}^2}\rho_m\Delta\ ,
		\end{equation}
		where
		\begin{equation}
			\frac{G_\text{eff}}{G} = 1+\mathcal{F}\ .
		\end{equation}
		
\bibliography{Biblio.bib}

\end{document}